# The 2020 Quantum Materials Roadmap

Abstract

In recent years, the notion of "Quantum Materials" has emerged as a powerful unifying concept across diverse fields of science and engineering, from condensed-matter and cold-atom physics to materials science and quantum computing. Beyond traditional quantum materials such as unconventional superconductors, heavy fermions, and multiferroics, the field has significantly expanded to encompass topological quantum matter, two-dimensional materials and their van der Waals heterostructures, Moiré materials, Floquet time crystals, as well as materials and devices for quantum computation with Majorana fermions. In this Roadmap collection we aim to capture a snapshot of the most recent developments in the field, and to identify outstanding challenges and emerging opportunities. The format of the Roadmap, whereby experts in each discipline share their viewpoint and articulate their vision for quantum materials, reflects the dynamic and multifaceted nature of this research area, and is meant to encourage exchanges and discussions across traditional disciplinary boundaries. It is our hope that this collective vision will contribute to sparking new fascinating questions and activities at the intersection of materials science, condensed matter physics, device engineering, and quantum information, and to shaping a clearer landscape of quantum materials science as a new frontier of interdisciplinary scientific inquiry.

**The 2020 Quantum Materials Roadmap – Table of Contents**

**Abstract**

# QUANTUM MATERIALS

## Feliciano Giustino and Stephan Roche

In recent years, the notion of "Quantum Materials" has emerged as a powerful unifying concept across diverse fields of science and engineering, from condensed-matter and cold-atom physics to materials science and quantum computing. Originally introduced to emphasize the exotic properties of unconventional superconductors, heavy-fermion systems, and multifunctional oxides, the definition of quantum materials has morphed into a much broader container that also encompasses research on topological properties, two-dimensional materials, and driven quantum systems. This convergence of diverse research areas is perhaps best exemplified by the coexistence of strong correlation, quantum criticality, and superconductivity in Moiré materials such as twisted bilayer graphene [1]. With this example in mind, it is natural to broadly define quantum materials as all those versatile materials platforms that allow us to explore emergent quantum phenomena as well as their potential uses in future technology.

Apart from unconventional superconductivity and Kondo physics, one of the earliest realizations of emergent phenomena in novel materials platforms can be traced back to the discovery of graphene. This discovery has opened a new dimension to explore unconventional transport properties of *massless Dirac fermions*, originally predicted to only occur at unreachable high energy scales [2].

The subsequent search for Dirac physics in materials beyond graphene has brought into focus the role of intrinsic spin-orbit coupling (SOC) effects, and led to the ground-breaking prediction of the existence of topological insulators and superconductors [3,4,5], as well as the ensuing experimental discovery of the former class [6,7]. Such new class of quantum materials is characterized by topologically protected massless Dirac surface states at the edges of two-dimensional materials or at the surfaces of their three-dimensional counterparts. These efforts were instrumental to the theoretical prediction and subsequent experimental discovery of Weyl fermions chalcogenide semimetals with their unique Fermi surface arcs [8,9]. These and many other advances have clearly shown that symmetry-based topological concepts are ubiquitous in materials physics, and offer unique opportunities to connect seemingly unrelated materials families and enable new discoveries.

In addition to its central role in topological quantum materials, SOC underpins a much wider array of phenomena in condensed matter. For example, in non-magnetic crystals with broken inversion symmetry, the Kramers degeneracy of electronic energy bands is lifted by SOC. This splitting arises from the interaction between the electron's spin and the Rashba field, that is the effective magnetic field that electrons experience in their rest frame when moving in an electric field. The Rashba field generates a wide variety of fascinating quantum phenomena, including the spin Hall effect (SHE), the spin–orbit torque (SOT), chiral magnons, skyrmions [10], and also Majorana fermions [11].

Moving beyond bulk three- and two-dimensional materials, tremendous advances in fabrication techniques have enabled the development of ultraclean (van der Waals) heterostructures based on atomically-thin two-dimensional crystals. In this new class of artificial materials, unique proximity effects that are absent in conventional materials lead to novel physical phenomena. Ground-breaking examples include topological order driving the quantum spin Hall effect in tungsten ditelluride [12], spin-charge conversion in graphene in proximity with transition-metal dichalcogenides [13], the demonstration of room temperature 2D ferromagnetism [14], the quantum anomalous effect [15,16], the observation of charge density waves and light-driven Floquet states [17], and the discovery of superconductivity in twisted bilayer graphene [18]. In addition, recent work on materials with strong SOC reported evidences of the long-sought after Majorana fermions. The Majorana particle is unique in high-energy physics insofar it is its own antiparticle [21,22]. Initial evidence for these particles has been obtained from the exploration of phenomena emerging at the boundary between strong spin-orbit coupling materials and superconductors, for example ballistic InSb nanowire devices and epitaxial Al-InAs nanowires [20]. Recent work also reported evidence of Majoranas in an iron selenium telluride superconductor [23]. Understanding these emerging quasiparticles is not only of fundamental interest, but it might also lead to applications, for example as the building blocks of non-abelian topological quantum computers [24].

The merging of strong correlations with topological matter creates yet another dimension for quantum materials, which might help in understanding and controlling complex phenomena such as high critical temperature superconductivity, the Kondo effect, and quantum criticality.

Taking a broader perspective, it is particularly remarkable that the nascent family of new quantum materials has ignited and accelerated efforts towards the development of next generation Quantum Technologies (QT). QT explicitly address individual quantum states and make use of the 'strangeness' of quantum physics such as superposition and entanglement, with the goal to build practical solutions for quantum computation, quantum sensing, and quantum metrology [25]. While quantum computers may still be some way in the future, quantum materials are already here to respond to the challenges posed by quantum computation, and to offer unique opportunities to imagine and realize the device architectures and computation paradigms that will be needed to achieve the so-called "quantum supremacy".

As the library of quantum materials grows larger, it is becoming increasingly important to be able to leverage modern data science techniques to catalogue, search, and design new quantum materials [26,27]. Machine learning (ML) and related artificial intelligence algorithms are expected to play a central role in this area. ML techniques can assist the high-throughput search for suitable materials for given applications, by harvesting the massive amount of raw data generated and stored by the computational materials science community. ML is particularly useful to identify patterns within large amounts of complex data, and many groups started to explore these methodologies for extracting knowledge and insight from materials databases [28]. ML is also seen as a powerful tool to advance the identification of quantum phase transitions and to deepen our understanding of many-body physics and quantum topological matter [25]. This nascent field holds great promise and is expected to play a fundamental role in progressing towards new materials solutions for next-generation quantum technologies.

In this vast and fast-moving research area, the aim of this Roadmap is to capture a snapshot of the latest developments in the field, and to offer a broad perspective on emerging horizons in quantum materials research. Although the field is already too big to be covered in all its aspects, we believe that many of the cornerstone materials driving innovation and underpinning emerging quantum phenomena are addressed in this collection.

It is often the case that radically new ideas emerge from scientific controversies and puzzling experimental data that challenge existing paradigms. It is our hope that, by pointing out such directions where further work and more in-depth analysis are called for, this roadmap will be able to stimulate new explorations and contribute to shaping this promising research field.

By giving voice to leading groups working on quantum materials from very different perspectives, we further expect that new fascinating questions will emerge at the intersection of materials science, condensed matter physics, device engineering, and quantum information, making quantum materials research a new frontier of scientific enquiry across disciplinary boundaries. We hope that our review will contribute to inspire much future work in the field.

## Complex oxides

Jin Hong Lee, Felix Trier and Manuel Bibes (Unité Mixte de Physique CNRS/Thales)

**Status**

After decades of work on bulk systems that led to the discovery of colossal magnetoresistance, multiferroicity and high-$T_C$ superconductivity, the field of complex oxides is now largely focused on thin film heterostructures. In such systems, materials with different properties can be assembled leading to cross-coupling between functionalities and giant responses. Often, interfaces between dissimilar materials exhibit unexpected physics arising from inversion-symmetry breaking and charge/spin/orbital reconstruction. Among a cornucopia of materials and physical effects explored in the last decade, three main families of quantum oxide systems have attracted a lot of attention: multiferroics, two-dimensional electron gases (2DEGs) and rare-earth nickelates $RNiO_3$ (R: rare-earth). Multiferroics simultaneously possess at least two ferroic orders and are usually antiferromagnetic and ferroelectric. Aside from having brought several new physical concepts such as improper ferroelectricity (induced by non-collinear spin order breaking inversion symmetry) or electromagnons (coupled lattice and spin excitations), multiferroics are attractive because the magnetoelectric coupling between spin and dipole orders allows for an electric field control of magnetism. Magnetoelectric coupling has fueled an intense research activity in single-phase materials such as $BiFeO_3$ [1] as well as in bilayers (combining a ferroelectric and a ferromagnet for instance) and in composites, with a view towards magnetoelectric sensors, memories or transistors. $BiFeO_3$, as well as some other ferroelectrics and multiferroics, also displays conductive domain walls between adjacent ferroelectric domains, which can be functionalized into nanoscale devices with rewritable circuitry.

Research on oxide 2DEGs was born with the discovery of a high-mobility metallic state at the interface between two insulators, $LaAlO_3$ and $SrTiO_3$ (STO)[2]. The 2DEG was later found to be superconducting and to possess a sizeable Rashba-type spin-orbit coupling (SOC)[3], both highly tunable by a gate voltage. A similar 2DEG can also form at the surface of STO[4] and by depositing a thin layer of a reactive metal such as Al on STO. Parallel to this research, huge progress was also made in oxide thin film synthesis, leading for instance to the observation of the integer and fractional quantum Hall effects in ultrapure ZnO/(Mg,Zn)O quantum wells[5] (Fig. 1b).

Another important direction in complex oxides has been the investigation of perovskite nickelate thin film heterostructures. Except for $LaNiO_3$, these compounds exhibit a metal-insulator transition at a temperature increasing with rare-earth atomic number. They also harbor a complex antiferromagnetic order at low temperature. Predictions of cuprate-like conductivity in carefully engineered nickelate superlattices[6] stimulated an intense research effort and superconductivity was finally found in nickelates in 2019, albeit through a different, chemistry-based approach to achieve an infinite-layer compound with similarities to cuprates[7] (Fig. 1c).

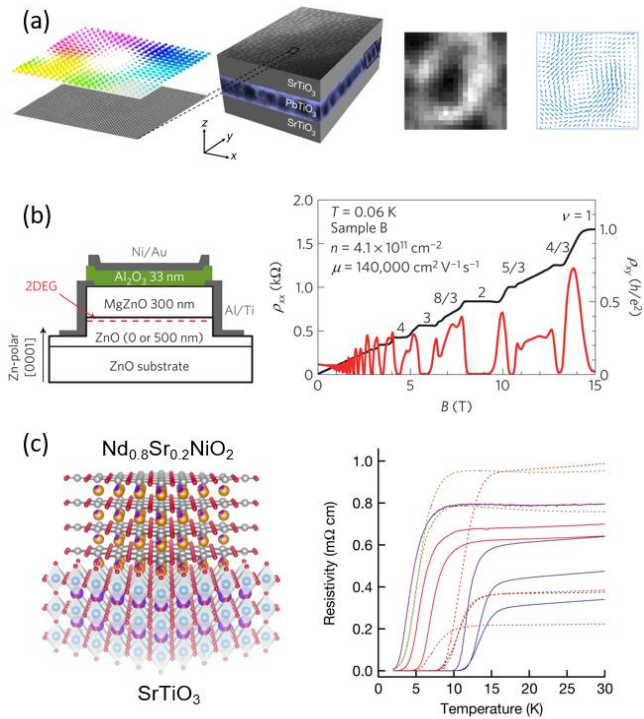

**Figure 1.** (a) Polar skyrmion structure observed in SrTiO₃/PbTiO₃ superlattices. The left panel shows a Ti-displacement vector map based on atomically resolved electron microscopy images. The rightmost images show annular dark field images of a skyrmion and a simulation [9]. (b) Fractional quantum Hall effect observed in a 2DEG at the ZnO/MgZnO interface [5]. (c) Superconductivity up to ~15 K in $Nd_{0.8}Sr_{0.2}NiO_2$ thin films on SrTiO₃ [7].

## Current and Future Challenges

Although complex oxides display a vast array of functionalities, most are restricted to low temperatures or their performance remains insufficient for room-temperature operation. Thus, an important challenge for the field is to design and realize materials with superior properties, eventually leading to applications. The field of multiferroics is particularly illustrative since BiFeO₃ has been and remains the only practical room-temperature multiferroic. Yet, its relatively weak magnetoelectric coupling complicates its integration into advanced devices such as the MESO transistor proposed by Intel[8] (MESO stands for MagnetoElectric Spin-Orbit, cf Fig. 2a), even though major progress was realized in the recent years[1]. The second leg of the MESO transistor is a spin-orbit element and the very efficient properties of spin-charge interconversion displayed by STO 2DEGs makes them interesting candidates for this application[3]. Yet, the room-temperature response still falls short by about one order of magnitude.

In terms of fundamental science, the emergence of chiral and topological spin and dipole textures[9] in magnetic oxides, ferroelectrics and multiferroics is opening an exciting new direction, with a view towards the electrical control of the generation and motion of such objects. Beyond STO 2DEGs, new systems with high mobilities have emerged such as stannates (e.g. La-doped BaSnO₃) or SrVO₃. SOC has also been introduced a key ingredient, adding to the usual spin, charge, lattice and orbital degrees of freedom, leading to exotic phases of matter, as in perovskite iridates. New superconducting phases have been discovered in ruthenates and more recently in nickelates[7], and much of their physics is still elusive. Despite various predictions and experimental attempts, oxide-based topological insulators have not yet been discovered.

The complex oxide community has also started to explore ultrafast phenomena to access transient states (including possible superconductivity at room temperature in cuprates[10]) or induce phase transitions, opening a vast new playground with many open questions.

**Advances in Science and Technology to Meet Challenges**

To address the challenges mentioned above a multifold strategy is required. Efforts must continue to design and synthesize new materials with high responses (e.g. a strong room-temperature multiferroic), looking beyond perovskites and towards phases that may be stabilized by epitaxy, post-growth chemical treatment, etc. We also have to identify materials combinations whose interfaces would exhibit new exotic properties (e.g. a 2DEG with Rashba SOC stronger than with STO); recent development with heavy-metal-based perovskites such as stannates and iridates are particularly promising in this direction. Another important aspect will be to synergetically exploit concurrent/competing degrees of freedom to realize novel states of matter (in the line of the ferroelectric skyrmions recently discovered cf Fig. 1a, but extended for instance to spin order in addition to dipole order). Combining perovskites with other quantum materials families such as transition metal dichalcogenides would also bring a vast array of possibilities in terms of new physics and new devices (particularly in light of the scarcity and relatively poor stability of ferromagnets and ferroelectrics in 2D materials family). Very recent progress in the preparation of complex oxide membranes will be an asset towards this goal and will also open possibilities for entirely new materials design rules inspired by the research on twisted graphene.

In terms of technological advances, efforts should continue to intertwine growth and characterization tools in ultra-high vacuum. *Operando* characterization is developing fast and in particular ARPES under piezostrain or electric field has been demonstrated. As previously mentioned, ps and sub-ps physics is under intense scrutiny and should develop further to reveal hidden phases and mechanism in complex oxide matter.

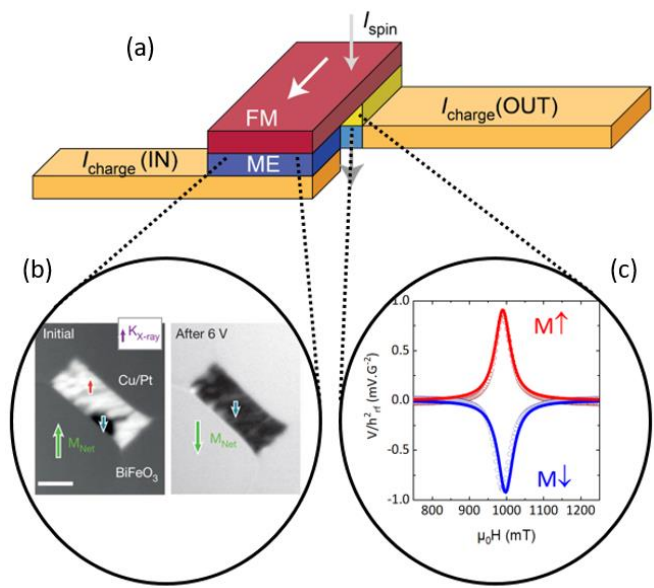

**Figure 2.** (a) MESO device proposed by Intel. Applying an input current charges the magnetoelectric capacitor (ME) and switches the magnetization direction of the ferromagnet (FM). Injecting a current into the FM produces a spin current that, when injected in the spin-orbit interface (yellow and blue bilayer), generates an output charge current through the inverse Edelstein effect [8]. (b) X-ray magnetic circular dichroism image of the ferromagnetic domain structure of a CoFe element deposited on top of a BiFeO$_3$ layer. Applying a voltage to the BiFeO$_3$ switches the CoFe magnetization [1]. (c) Voltage generated by the conversion of a spin current into a charge current by a AlOx/STO 2DEG through the inverse Edelstein effect. When the magnetization of the FM injecting the spin current is reversed, the charge current produced changes sign [3].

**Concluding Remarks**

In summary, the field of oxide-based quantum matter is moving at a speedy pace, which has led to several important discoveries in the most recent years (ferroelectric skyrmions, giant spin-charge conversion, superconductivity in nickelates, etc). This opens new research directions but also poses many physics questions on the conditions required to stabilize such new electronic states, the underlying mechanisms, the associated atomic, electronic and magnetic structure, etc. However, in addition to advancing the understanding the physics of complex oxides – that has paved research in condensed matter for the last 50 years – the field must also move faster towards applications (beyond the use of e.g. ferroelectrics and piezoelectrics as infrared detectors, sensors, micro-actuators or electro-optic components, all relying on bulk systems and decades-old physics). The MESO transistor recently proposed by Intel[8] and operating on multiferroics and spin-orbit materials is not only one fantastic opportunity for the field but also a broader sign that major information and communication technology companies are now ready to consider exotic materials such as complex oxides for beyond CMOS solutions.

**Acknowledgements**

Financial support from ERC Consolidator Grant 615759 "MINT" and ERC Advanced Grant 833976 "FRESCO" is acknowledged.

# Quantum Spin Liquids

Stephen M Winter[1] and Roser Valentí[1]

1 Institut für Theoretische Physik, Goethe-Universität Frankfurt, Max-von-Laue-Strasse 1, 60438 Frankfurt am Main, Germany

*Status.* Quantum materials research focusses on phases of matter in which uniquely non-classical phenomena, such as quantum fluctuations, entanglement and quantized topological numbers play an essential role in establishing physical properties. Quantum spin-liquids (QSLs) represent quintessential examples of such phases [1], in which the elementary degrees of freedom are the magnetic spins (qubits). Provided symmetry-breaking orders can be avoided by careful tuning of the interactions, such spins may adopt macroscopically entangled ground states, featuring e.g. topological orders and magnetic excitations with unconventional statistics. These phases are broadly referred to as QSLs.

At present, spin models represent minimal models for a wide variety of desirable phases of fundamental interest and for applications in e.g. quantum information sciences. The field spans a broad range of topics and goals, from quantum information to crystal engineering. A significant focus continues to be the unambiguous identification of a particular QSL phase in a real material, which requires: (i) theoretical classification of possible QSLs, (ii) motivated materials searches, (iii) a variety of experimental tools, and (iv) interpretation of experiments aided by numerical and first-principles simulations (see Figure 1).

Experimental criteria in the search for QSL materials often focusses on identifying what they are not, i.e. systems where conventional order is avoided down to lowest measurable temperatures. Designing novel experiments that instead directly probe their unconventional excitations or entanglement structure, continues to be a necessary target.

Traditional candidates for QSL materials are geometrically frustrated networks of spins [2] such as Kagome Herbertsmithite or triangular lattice of κ-(*ET*)$_2$Cu$_2$(CN)$_3$. However, recent years have seen an increased interest in heavy *d*- and *f*-block elements for which strong spin-orbit coupling induces highly anisotropic magnetic couplings. Examples include, but are not limited to, rare-earth pyrochlores [3] and various Kitaev-QSL candidates [4,5]. These materials feature additional fluctuations driven by competition between non-commuting interactions on different bonds. However, the departure from spin-rotational invariance (and, in some cases, spin > 1/2) implies a large number of independent coupling constants are required to completely parameterize their Hamiltonians. This places increased importance on the application of a variety of experimental and *ab-initio* approaches to locate given materials within broader phase diagrams.

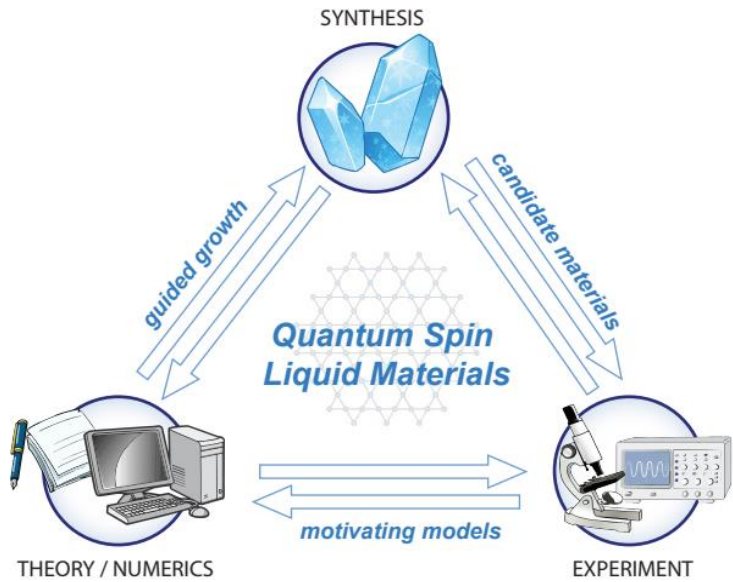

**Figure 1.** Schematic representation of the research cycle for QSL materials.

*Current and Future Challenges.*
In the discovery, characterization, and utilization of QSL materials, several challenges appear. Here, we highlight a few key aspects:

- By their nature, QSLs often occur in the vicinity of many competing phases, where relevant energy scales are artificially reduced due to frustration. This increases the importance of understanding sub-dominant effects - such as disorder - which may yield novel aspects and have a dramatic impact on experiments and the stability of given ground states [6,7].

- Due to the complexity of magnetic Hamiltonians in modern QSL materials, estimation of the many sub-dominant couplings via experiment alone is often unfeasible; state-of-the-art *ab initio* methods are required. These methods must go beyond density-functional theory, incorporating explicit many-body treatment of spin multiplets and spin-orbit coupling in order to provide access to all symmetry-allowed couplings.

- In order to aid in the identification and classification of QSL materials, there is need for "smoking-gun" probes of particular QSL states, as well as general experimental methodologies to probe non-local entanglement.

- Continued development of numerical methods - applicable to arbitrary quantum spin models - is needed to predict properties (particularly excitation spectra), in order to compare with experiments and characterize models.

- Regarding model development: It is generally accepted that the low-energy theory of many QSLs features emergent gauge symmetries and corresponding collective excitations. The discovery of spin models with exact solutions having a $Z_2$ gauge structure (such as the Kitaev's honeycomb model [4,5]) have bolstered this view. A challenging task is to find exactly solvable models for other phases such as U(1) spin-liquids, which have been proposed for e.g. the Kagome systems and triangular lattice organics.

☐ Even when exactly solvable models are found, they often have interactions that are unrealistic for real materials. Identification of realistic spin models exhibiting given ground states – and microscopic conditions for engineering them in real materials - therefore continues to be a significant focus.

☐ Finally, in the continued search for new QSL materials, a better understanding of how to tune materials towards desired parameter regimes is of continued interest. Various approaches are being explored, including combinations of chemical crystal engineering, physical pressure or strain, and exfoliation of 2D materials.

*Advances in Science and Technology to Meet Challenges.*

From the experimental perspective, development of novel "smoking-gun" measurements is a crucial consideration. For example, the observation of quantized thermal Hall response in a magnetic insulator represents a nearly unique signature of a chiral QSL, although better understanding of the role of phonons in these experiments is important for their interpretation [8]. Similarly, non-linear spectroscopies probing higher order spin correlation functions, such as RIXS, may provide more direct probes of specific excitations in QSLs if resolution can be continuously improved. To provide a measurement of non-local entanglement, spectroscopies employing multiple entangled photons is also an interesting future consideration. Similarly, the realization of QSL-equivalent states in cold-atom experiments may provide access to additional modes of measurement.

From the perspective of new material searches, finer chemical control of disorder is an ever-present target. For understanding the effects of disorder, strain, pressure, further development of *ab initio* structure predictions is likely to play an increasingly prominent role.

From the modeling perspective, adapting various numerical many-body methods to treat the lower-symmetry interactions in $d$- and $f$-block materials is highly desirable. For the calculation of dynamical correlation functions, the extension of traditional ground-state methods has proved valuable, e.g. time-dependent DMRG and VMC [9]. Advancements in the efficient representations of ground- and excited-state wavefunctions in 2D and 3D models, through e.g. tensor-network states, or machine-learning inspired forms [10], will no-doubt aid these approaches.

Finally, one of the motivations in the search for QSLs is the utilization of their non-trivial excitations to perform quantum computations or store quantum information. If such technologies are to be realized, they will eventually require protocols to experimentally create and manipulate such excitations.

*Concluding Remarks*

QSLs are very exciting and, at the same time, elusive states of matter. In particular, the explosion of interest in topological quantum matter has coincided with the discovery of many spin models of fundamental interest, from Kitaev's Toric code, to fracton X-Cube models. While these models serve as an inspiration, conclusive experimental demonstrations of many predicted phases remain rare. This elusivity is likely due to a combination of (i) rarity of QSL phases in realistic phase diagrams, (ii) sensitivity of QSLs to sub-dominant perturbations such as disorder, and (iii) lack of "smoking-gun" experiments to uniquely identify the ground state of candidate materials. Given significant and steady progress over the years – from the conception of QSLs, to the discovery of realistic candidate materials – it seems inevitable that these challenges will be met.

## Acknowledgements
We acknowledge financial support by the Deutsche Forschungs-gemeinschaft through Grant VA117/15-1. This research was supported in part by the National Science Foundation under Grant No. NSF PHY-1748958

**Twisted two-dimensional layered crystals**

Young-Woo Son, Korea Institute for Advanced Study, Seoul 02455, Korea

**Status**

When repetitive lattice structures are overlaid against each other, a new superimposed periodic lattice, called as the moiré pattern, emerges. Recent progress in stacking two-dimensional materials (2DMs) enables the patterns to occur at the atomic scale. Among various possible atomic scale moiré lattices with 2DMs, twisted bilayer graphene (TBG) shown in Fig. 1(a), a single layer graphene on top of the other with a rotational stacking fault, has attracted a considerable attention. The size of moiré unit cell of TBG is inversely proportional to the twist angle [Fig. 1(b)] and the Fermi velocity at the Dirac point decreases with lowering the twist angle. Specially, for a certain set of angles near 1 degree called as magic angles, the velocity is strongly reduced, realizing flat bands [1]. A set of experiments [2] exploring the flat bands in TBGs with a magic angle has ignited worldwide research activities in pursuit of new state matters.

The magic angle TBGs (MATBGs) are insulating thanks to Mott insulating mechanism while a transition to a superconducing state is triggered under appropriate doping as shown in Fig. 2(a) [2]. This seemingly similar phase diagram [Fig. 2(b)] with high temperature superconductors (HTCSs) poses a significant challenge to understand correlated physics related with doped Mott insulators in low dimension [2]. Unlike HTCSs, the MATBGs can be easily doped without chemical means and be examined with various direct local probes on their bare surfaces [3]. With MATBGs, the other collective phenomena such as ferromagnetisms [4] as well as topological properties [5] have been discovered. Besides TBGs, trilayer graphene [6] and double bilayer graphene [7] with twist angles have also demonstrated their unique potentials in realizing novel quantum states.

Besides correlated physics in TBGs with small angles, there are also noteworthy physics in TBGs with large twist angles. One is a realization of effective 12-fold quasicrystalline order in TBGs with 30 degrees twist angle [8] and the other is a higher-order topological insulating phase in commensurate TBGs with large twisted angles [9]. These works demonstrate that TBG with an appropriate twist angle can be a versatile platform for many interesting concepts in the 21$^{st}$ century condensed matter physics.

Together with graphitic materials, several new proposals and experiments for twisted geometries with other 2DMs such as transition metal dichalcogenides (TMDCs) [10] have been put forward, enriching the playground of flat bands systems.

**Current and Future Challenges**

Understanding the nature of insulating and superconducting phases in MATBGs is one of the important challenges in the field. The linear enhancement of resistivity in a broad range of temperature may indicate a 'strange metal' phase discussed in correlated materials [11] although the issue remains controversial [12] and needs to be clarified. To uncover a true nature of MATBGs, it is important to describe electronic states, phonon dispersions and their interactions theoretically. Models and calculations for electronic structures [13], phonon dispersions [14] and their mutual interactions [12,15] have been established quickly. However, to capture correlated physics, it has become crucial to incorporate strong interactions in future efforts.

Considering various degrees of freedom in manipulating TBGs and other twisted 2DMs, the new physics arising from quasiparticles and collective phenomena seem to be enormously diverse here. In particular, the complete quenching of kinetic energy of electrons reveals its consequences only in few physical phenomena such as superconducting phases [2,6,7] and anomalous quantum Hall effects [5] to name a few. So, more correlated topological phases can be expected. As already shown theoretically for spin-triplet superconductor in trilayer graphene [6], highly nontrivial superconducting and topological phases may be realizable. Since we can control the ratio between kinetic and Coulomb interaction by varying the twist angle, the twisted system could be one of textbook experiment tools for studying competing interactions in low dimensions. With other controlling knobs such as external electromagnetic fields and pressure, we also expect various strongly correlated optoelectric physics in MATBGs and in twisted TMDCs as already evidenced in simple TMDCs. Heterogeneously stacked twisted 2DMs, e.g., Boron-Nitride layers with TBG, have already demonstrated its unique merits [5,6] so that we expect more diverse phenomena from twisted 2DM with different physical and chemical combinations.

The effective quasicrystalline order realized in a TBG with twist angle of 30 degrees also poses an important challenge in studying the relationship between lattice ordering and electron localization. Since the graphene quasicrystal displays strong localization far away from its Fermi energy [8], either large amount of doping or other 2DM quasicrystals may realize the system to look for critical physics for electron localization related with low dimensional quasicrystals.

**Advances in Science and Technology to Meet Challenges**

Since the most MATBGs with rotational stacking faults are hitherto fabricated through a 'tear and stick' method [2], a controllable way of placing layers with a desired twist angle is required. Some progresses have been made [16] based on nanomechanical or robotic technologies. However, considering important roles of inhomogeneous twist angles in MATBGs [17], more reliable and facile methods with high device yield should be developed. Together with these specific issues for twisted structures, other technological challenges that are common to all 2DM researches such as interfaces between 2DM and three-dimensional materials also should be improved.

One of the most interesting physics related with twisted geometries of homogeneously or heterogeneously stacked 2DMs are related with nearly flat bands [1,2,10]. So, the central understanding of novel physics relies on the analysis of the Hubbard model in two-dimension where the strong local Coulomb interaction dominates over the kinetic energy between sites in the lattice. Almost all twisted geometries involve with two or three hexagonal lattices so that the resulting moiré superlattices realize the Hubbard model on triangular symmetry. Current models from maximally localized Wannier orbitals constructed on flat bands point to interesting non-local or extended Hubbard Coulomb interactions and multiplet structures [13]. A few experiments e.g. Ref. [3] already reveal an anomalous charge ordering related with the triangular Hubbard model. Therefore, precise theoretical and experimental understandings on this model with and without other quasiparticles such as phonon in TBGs [14,15] will provide a way to realize several new phases with distinct topological properties [3-7].

Considering a vast expansion of available 2DMs with different characteristics, the current set of examples for twisted 2DMs may be the tip of iceberg [10]. To explore almost limitless combinations for twisted heterostructures, experimental techniques for a precise angle control between different 2DMs are crucially need, together with development of the theoretical and computational machineries to describe interlayer interactions between incommensurate heterogenous lattices and the more sophisticated low energy theories based on them. Being similar with small angle twisted 2DMs, for quasicrystalline ordering, the precise angle of 30 degrees between the hexagonal layers are required so that practical techniques of building systems onto suitable substrates need to be developed beyond currently highly complicated approaches which limit the fabrication to very few devices [8].

**Concluding Remarks**

The field of physics and materials sciences on twisted layered structures of 2DMs is very rapidly growing. It already provides us intriguing parallelism between phase diagrams of MATBGs and HTCSs. They also demonstrated several novel emergent collective phenomena of combined systems that are not characteristics of individual layers. So, the twisted 2DMs will soon become a new viable system for strongly correlated physics in low dimensions. Considering its very early stages in developments, we expect that the field will diversify itself by using various different 2DMs and that related science and technology for both fundamental researches and technological applications will thrive in very near future.

**Acknowledgements**

YWS was supported by the NRF of Korea (Grant No. 2017R1A5A1014862, SRC program: vdWMRC Center) and by KIAS individual grant (CG031509).

robotic searching and assembly of two-dimensional crystals to build van der Waals superlattices *Nat. Comm.* **9** 1413

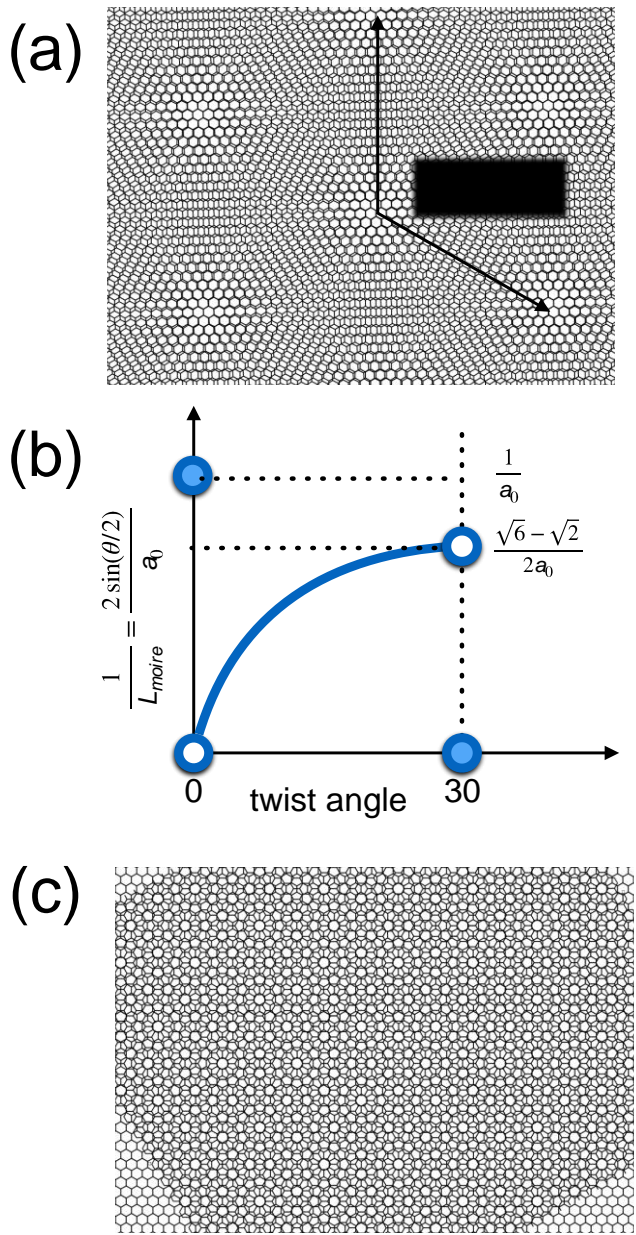

**Figure 1.** (a) Schematic atomic lattice structures of bilayer graphene with a rotational stacking fault. Here, a twist angle between top and bottom layer is 3 degrees. Moiré lattice with a moiré unit vector of $L_{moiré}$ much larger than graphene atomic lattice of $a_0$ drawn with arrows. (b) A relationship between twist angle ($\theta$) and $1/L_{moiré} = 1/|L_{moiré}| = 2\sin(\theta/2)/a_0$. The moiré lattice vectors do not need to match the actual commensurate lattice constants for twisted system always. In case of $\theta = 0$, the size of the vector diverges (shown in empty circle) while the actual periodicity is $a_0$ (full circle). At 30 degrees, a moiré lattice of its unit vector size of $(\sqrt{6}+\sqrt{2})a_0/2$ exists (empty circle) although the system has no translational symmetry, or a diverging unitcell (empty circle). (c) The 12-fold quasicrystalline order in twisted bilayer graphene with a twist angle of 30 degrees.

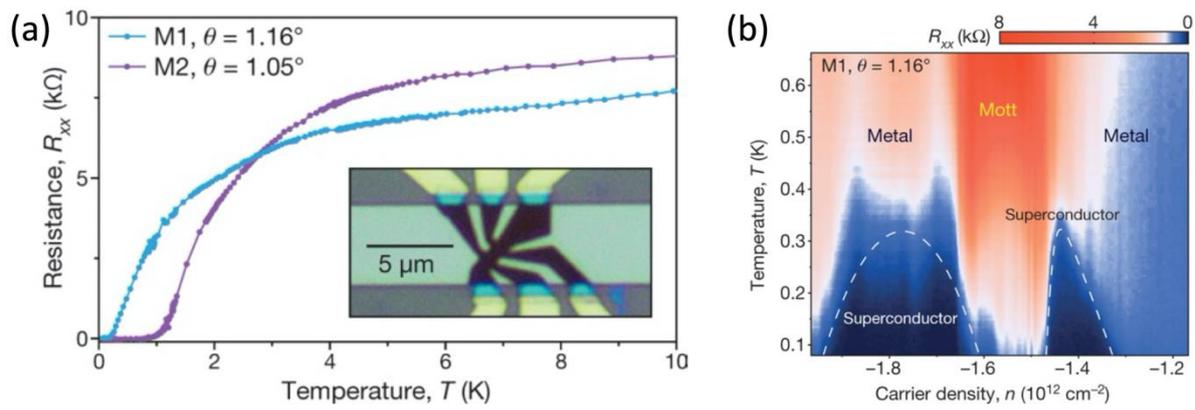

**Figure 2.** (a) Four-probe longitudinal resistance ($R_{xx}$) measurement for two different MATBGs (reproduced from [2]). The both samples show superconducting transitions at 0.5 K and 1.7 K, respectively with small doping. (b) A phase diagram of $R_{xx}$ of a MATBG versus temperature near the charge neutral point of $n = -1.58 \times 10^{12}$ cm$^{-2}$ (reproduced from [2]). Two superconducting domes are located above and below the charge neutral Mott insulating state.

# Cuprate superconductors

Louis Taillefer[1,2]

*1 Département de physique, Institut quantique, and RQMP, Université de Sherbrooke, Sherbrooke, Québec, Canada*

*2 Canadian Institute for Advanced Research, Toronto, Ontario, Canada*

## STATUS

Superconductivity in copper oxides called "cuprates" was discovered in 1986. In the three decades that followed, cuprates emerged as perhaps the most remarkable and enigmatic quantum materials, with striking discoveries being made up to this day. In the first decade (1986-1997), the intensive search through hundreds of different ways to stack $CuO_2$ planes together with intervening layers led to a rapid increase in the superconducting transition temperature $T_c$, which reached its current maximal value of 133 K in 1993 – still the record today for the highest $T_c$ at ambient pressure. Two major achievements of this first decade are the conclusive demonstration that the superconducting order parameter of cuprates has $d$-wave symmetry[1,2], and the discovery of "stripe order", a density-wave state of intertwined spin and charge modulations[3], that coexists and competes with superconductivity.

In the second decade (1998-2008), it became clear that at the centre of the cuprate phase diagram lies a mysterious phase called "the pseudogap phase" (Fig. 1), whose properties have since been the focus of intense scrutiny. As the temperature is reduced below the pseudogap temperature $T^*$, a partial gap opens in the spectral weight for wavevectors $(\pm \pi, 0)$ and $(0, \pm \pi)$ – as detected by angle-resolved photoemission (ARPES)[4], for example – and the density of states drops dramatically – as detected in the specific heat[5], for example. The nature of the pseudogap phase is still a matter of active debate today. Numerical solutions of the Hubbard model obtain a pseudogap from strong correlations without breaking any symmetry[6], but there are proposals for a state that does break symmetry – for example, a state of current loop order that breaks time-reversal but not translational symmetry[7].

A fundamental question is : what is the Fermi surface in the pseudogap phase (at $T = 0$)? Outside the pseudogap phase, at high doping ($p > p*$), angle-dependent magneto-resistance (ADMR)[8], ARPES[9] and quantum oscillations[10] in Tl2201 with $p \sim 0.3$ all detect a large hole-like cylinder containing $n = 1 + p$ holes per Cu, consistent with band structure calculations. Inside the pseudogap phase ($p < p*$), ARPES sees a breaking up of that large Fermi surface, with only disconnected Fermi arcs remaining in the nodal ($\pm \pi, \pm \pi$) directions of the Brillouin zone[11]. By contrast, quantum oscillations detected in YBCO at $p \sim 0.1$ reveal a coherent and closed Fermi surface at low temperature ($T < 10$ K), once superconductivity has been removed by application of a magnetic field[12].

In the third decade (2009-2019), charge order was discovered in YBCO by NMR[13] and then observed by X-ray diffraction[14,15], but without spin order, showing that spin and charge modulations do not necessarily come together. The additional periodicity associated with charge order, which breaks translational symmetry and therefore reconstructs the Fermi surface, provides a natural mechanism for the electron-like Fermi pocket seen in quantum oscillations over the same doping range ($0.08 < p < 0.16$). The fact that this range ends before the critical doping $p*$ at which the pseudogap phase ends shows that the two phases are distinct. High-field Hall measurements reveal a rapid change in the carrier density, from $n = 1 + p$ above $p*$ to $n = p$ below $p*$ (refs. 16, 17). STM measurements using quasiparticle interference deduce a Fermi surface that goes abruptly from arcs below $p*$ to a full circle above $p*$ (ref. 18). Whether the actual Fermi surface just below $p*$ – in the pure pseudogap phase without charge order – consists in fact of small closed hole pockets at nodal positions that contain $p$ holes per Cu is a key open question.

Just above $p^*$, the resistivity of cuprates exhibits a strikingly linear temperature dependence down to the lowest temperatures[19,20]. The microscopic mechanism responsible for this phenomenon, also observed in heavy-fermion metals, pnictide and organic superconductors[21], for example, stands today as one of the major puzzles in condensed-matter physics.

## CURRENT AND FUTURE CHALLENGES

In 2020, several challenges are presented to us by cuprate superconductors[22,23]. The obvious practical one is how to increase $T_c$. A second challenge is the perfect $T$-linear resistivity, seen in what are sometimes called "strange metals" – because they do not exhibit the standard $T^2$ dependence expected of Fermi liquid theory. It was recently shown that the slope of the $T$-linear dependence in cuprates satisfies the so-called "Planckian limit" (ref. 24), whereby the inelastic scattering time is given by Planck's constant divided by thermal energy, *i.e.* $\tau = \hbar \, / \, k_B \, T$ – as indeed it does in other materials[25]. How does this Planckian dissipation work?

A third and broad challenge is to elucidate the nature of the pseudogap phase. A first aspect is its critical point at $p^*$. Whereas most physical properties of cuprates exhibit a smooth crossover as a function of temperature (through $T^*$), the change in carrier density – or in the conductivity[26] – as a function of doping (through $p^*$) is rather sharp. Is this a quantum critical point, akin to that seen in organic[27], heavy-fermion[28] or pnictide[29] superconductors? In the phase diagrams of these three families of materials, a phase of long-range antiferromagnetic order is suppressed by some tuning parameter like pressure or concentration and a dome of superconductivity forms around the quantum critical point – suggesting a common pairing mechanism[30]. In cuprates, recent specific heat measurements find the same thermodynamic signatures of quantum criticality as those seen in heavy-fermion and pnictide superconductors[31]. But in cuprates, there is no long-range antiferromagnetic order near $p^*$! Is it possible that short-range magnetic order is sufficient to produce the same signatures? If translational symmetry is not broken on a long range, does the change of carrier density necessarily imply a topological order[32] ? Could the critical point be weakly first-order, as obtained in some Hubbard model calculations[33] ?

A second aspect is the Fermi surface associated with the low carrier density $n = p$ below $p^*$. What is it? A third aspect is the various indications of broken symmetries, such as time reversal[34], inversion[35], and rotation[36,37,38]. How can we make sense of these?

## ADVANCES

The maximal $T_c$ in cuprates has not moved since 1993, and insights from new materials would be useful. In 2008, the discovery of superconductivity in iron-based superconductors certainly showed that a high $T_c$ was possible in other materials (up to 55 K) – the pairing in this case being most likely associated with antiferromagnetic spin fluctuations. A possible avenue of investigation is the role of phonons, which could further enhance $T_c$, as proposed for monolayer FeSe films[39]. In 2018, the striking discovery of superconductivity in twisted bilayer graphene[40] provides an entirely different context for pairing, with $T_c$ values even higher than in cuprates when measured relative to the Fermi temperature. Another relevant discovery, made in 2019, is superconductivity in nickel oxides[41]. The comparison of cuprates to these three different superconductors is sure to be instructive.

Progress on the subject of Planckian dissipation would require having more direct measurements of the scattering time – for example via optical studies or ADMR. The recent report of $T$-linear resistivity with a Planckian slope in twisted bilayer graphene[42] provides a great platform for controlled studies of this phenomenon.

For the pseudogap phase, one of the most decisive measurements would be to see quantum oscillations in the pure phase (without charge order). In the cleanest materials like YBCO, Hg1201 and Tl2201, all with a high maximal $T_c$ (~90 K), the magnetic fields necessary to suppress superconductivity down to $T \sim 1$ K are in excess of 100 T (ref. 43) – requiring new techniques for detection in ultra-high fields. Seeing oscillations with a frequency $F = p\, \Phi_0 / 2$ – coming most naturally from nodal hole pockets with a total carrier density $p$ – would go a long way to establishing the wavevector $(\pi, \pi)$ as a fundamental organizing principle of the pseudogap phase.

Further advances in computational methods would help tighten the connection between the Hubbard model and the real cuprates – ideally by being able to resolve the expected Fermi surface in the pseudogap phase – and other theories of the pseudogap phase – for example the SU(2) gauge theory of fluctuating antiferromagnetism, with its topologically ordered Higgs phase[44].

The recent observation of a large thermal Hall effect in the pseudogap phase of cuprates[45] raises an entirely new question: by what mechanism can the pseudogap phase confer chirality to neutral excitations? The same question is now being asked in the context of spin liquids, insulating materials where magnetic order is frustrated by the structure, in favor of new states of quantum matter that are expected to potentially harbor new excitations such as spinons or Majorana modes, whose presence may be detectable by the thermal Hall effect[46].

**CONCLUDING REMARKS**

A promising roadmap for tackling the biggest questions about cuprate superconductors – maximal $T_c$, Planckian dissipation, pseudogap phase – is to explore connections with other materials, in particular iron-based and nickelate superconductors, twisted bilayer graphene and spin liquids.

**ACKNOWLEDGEMENTS**

L.T. acknowledges support from the Canadian Institute for Advanced Research as a CIFAR Fellow of its Quantum Materials Program.

**FIGURE**

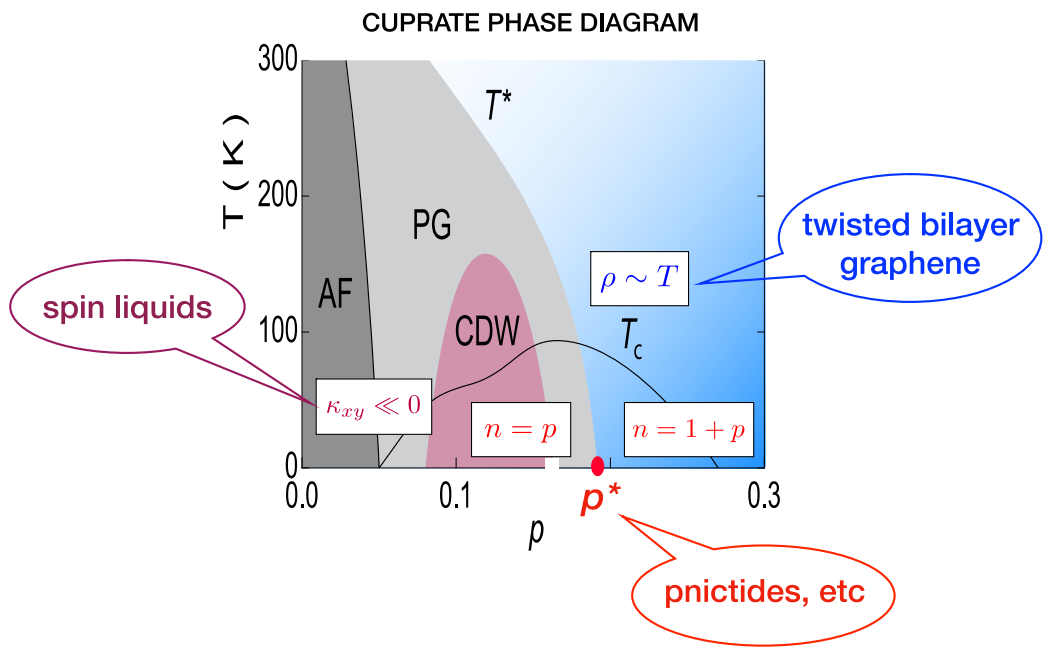

**Fig. 1 | Phase diagram of cuprate superconductors.**

Temperature-doping phase diagram of cuprate superconductors, showing the phase of superconductivity below $T_c$ (dashed line), the pseudogap phase below $T^*$ (PG, grey), the phase of long-range antiferromagnetic order below TN (AF, dark grey) and the region where charge-density-wave (CDW) order exists (purple), centered around $p \sim 0.12$. The pseudogap phase ends at the critical doping $p^*$ (red dot). When superconductivity is suppressed with a magnetic field, various signatures of this critical point are revealed, including: $T$-linear resistivity down to $T = 0$ at $p = p^*$; drop in carrier density from $n = 1 + p$ above $p^*$ to $n = p$ below $p^*$; large negative thermal Hall conductivity $\kappa_{xy}$ at low $T$ below $p^*$, down to $p = 0$. In the foreseeable future, different families of materials are likely to shed new light on some of the major puzzles of cuprates, in particular *spin liquids* on the nature of the enigmatic pseudogap phase, *twisted bilayer graphene* on the intriguing Planckian dissipation, and *pnictide and nickelate superconductors* on the mechanism for superconductivity.

**Ultrathin layered superconductors**

Christoph Heil, Institute of Theoretical and Computational Physics, Graz University of Technology, NAWI Graz, 8010 Graz, Austria

**Status**

This section is going to review recent advances and current challenges with respect to atomically thin superconductors, placing particular focus on transition metal dichalcogenides (TMDs), which are chemically and physically very diverse, thus exhibiting very rich low temperature phase diagrams.

Superconductivity in two dimensions (2D) has fascinated the scientific community since the late 1930's, but due to the many surfaces and interfaces of a 2D system resulting in the tendency for structural and chemical disorder, early experimental work was restricted to the investigation of amorphous or granular thin films of metals and alloys that nonetheless allowed the study of important properties of the superconducting state independent of the crystal orientation and structure. It was found, for example, that the angular dependence of the upper critical magnetic field exhibits a peak-like response if the magnetic field is parallel to the thin film, which has subsequently been explained by a phenomenological theory based on the Ginzburg-Landau model and is considered a signature feature of 2D superconductivity [1].

Recent technological advances, such as molecular beam epitaxy (MBE), pulsed laser deposition, electric double layer transistors (EDLT), and mechanical exfoliation allow the fabrication of highly crystalline and atomically-thin samples, and experimental progress with techniques such as scanning tunnelling spectroscopy and angle-resolved photoemission spectroscopy enable researchers to investigate materials' properties that are intimately linked to their crystal structure. On the theoretical side, the availability of high-performance computing clusters paired with the advances made on describing a material *ab-initio* enables researchers to quantitatively reproduce and even predict quantities like the superconducting critical temperature $T_c$.

Particular efforts were made to shed light on quantum phase transitions, as for example between the insulating and superconducting state, or between metallic charge-density wave (CDW) phases and superconductivity, as well as Ising-, interface-, and topological superconductivity [1, 2]. Of particular interest in this respect are TMDs, layered materials whose layers are weakly held together by van-der-Waals forces. This allows the exfoliation down to the monolayer limit, making them one of the thinnest superconductors made to date (Fig. 1).

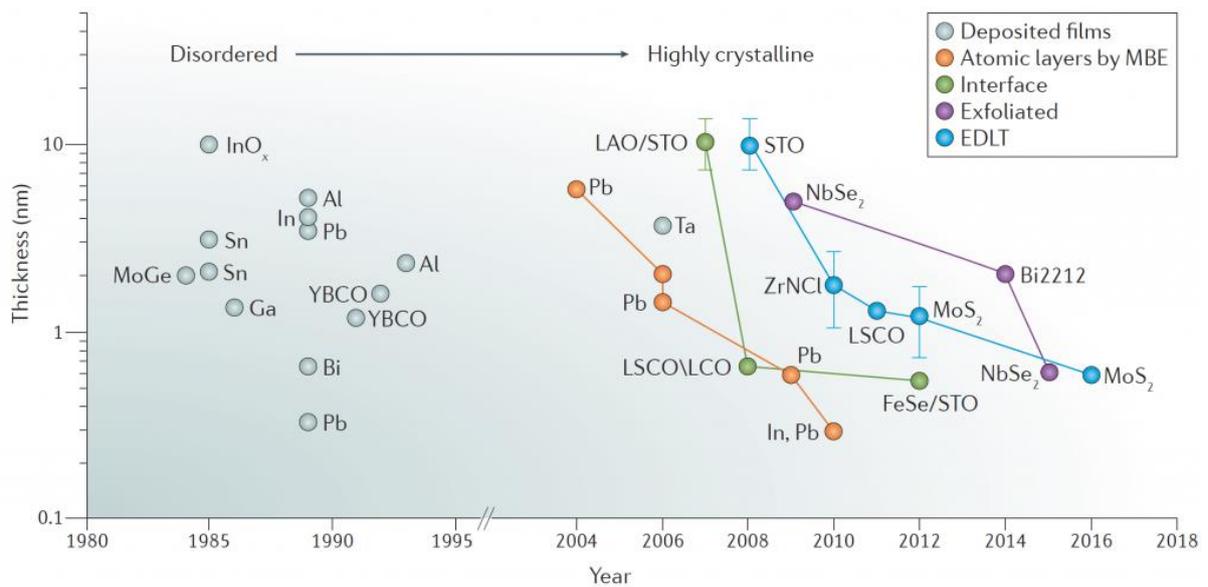

**Figure 1.** Evolution of the thickness of ultrathin superconductors since 1980. Reprinted by permission from [1].

## Current and Future Challenges

TMDs have attracted great research interest as their superconducting phase often coexists with CDW phases. $NbSe_2$, for example, exhibits a CDW phase below 33K and is superconducting with a $T_c$ ~7K in the bulk. When thinned down, $T_c$ decreases monotonously reaching a $T_c$ ~2 K in the single-layer limit (Fig. 2, left) [1-3]. The existence of a CDW for the monolayer at a higher $T_{CDW}$ than in the bulk has been predicted theoretically [4], and while a CDW has been observed experimentally, studies are in disagreement at which temperature $T_{CDW}$ this phase sets in. Interestingly, a recent theoretical study predicts that both $T_c$ and $T_{CDW}$ increase in the closely related $TaSe_2$ when thinned down from bulk to monolayer [5]. Experimentally, a strengthened CDW and increasing superconductivity were observed in few-layer samples. If SC survives in the monolayer limit of $TaSe_2$, and why SC reacts differently in different TMD materials when thinned down are just two questions that need to be investigated further.

One of the most intriguing features of SC in ultrathin TMDs is the fact that it survives huge in-plane magnetic fields, factors of 5 and larger than the corresponding Pauli paramagnetic limit (Fig. 2, right) [3,6]. A possible explanation for this behaviour is based on strong spin-orbit interactions leading to almost perfect spin splitting in certain valleys of the Brillouin zone. In the superconducting phase, the Cooper pairs are formed between these valleys, resulting in an internal magnetic field protecting SC [1, 3]. A recent study extended this concept of *Ising superconductivity* to certain centrosymmetric 2D materials, opening up a very interesting path for further research [7].

Within the versatile class of TMDs one can also find topological insulators. Monolayer $WTe_2$, a quantum spin Hall insulator, can be reversibly tuned to become superconducting using the electrostatic field effect [8]. The proximity of a topological insulating state to a superconducting phase offers many new and interesting research directions. For one, it has not yet been clarified whether the superconducting state itself is topologically nontrivial or not, and it would be fascinating to investigate Majorana modes within these materials.

## Advances in Science and Technology to Meet Challenges

Recent theoretical improvements allow the description of the superconducting phase fully *ab-initio* and to look at quantities like the superconducting gap function or the electron-phonon coupling band-and wave-vector resolved, offering unprecedented insight into the origins of SC in TMDs [9]. To improve the theoretical description further, it would be important to include anharmonic and non-adiabatic effects, as the superconducting phase in these materials appears in close vicinity to quantum phase transitions and lattice instabilities [4, 5, 9]. Also, in order to be able to describe concepts like Ising superconductivity, spin-polarization needs to be taken into account in the calculations.

With respect to superconductivity in the vicinity of nontrivial topology, fascinating new phases and states might yet to be unravelled, possibly requiring the consideration of new models to describe SC there. A recent theoretical work, for example, proposes a way to realize Majorana fermions in monolayer $NbSe_2$ and $TaS_2$, employing magnetic fields higher than the Pauli limit, but lower than the in-plane critical field to turn them into nodal topological superconductors [10].

At this point it should also be noted that in most experimental setups, the ultrathin TMD samples are placed on a substrate and protected from atmospheric influences by a capping layer, thereby representing a complex layered heterostructure. Great care has to be taken to attribute any measured physical quantity or phenomenon to its origin within this heterostructure. On the other hand, a lot of theoretical work is based on freestanding mono- and few-layers to keep computational costs to a minimum. While this allows for investigating effects confined to the ultrathin layer, it does not quite represent most experimental setups, as interfaces can have important influences on quantities like electron-phonon interaction and superconductivity. One of the key advances in the next few years should therefore be to improve our understanding of interlayer physics with respect to quantum phase transitions and superconductivity in particular. On the theoretical side, this means to consider large supercells and heterostructures to better correspond to experimental samples.

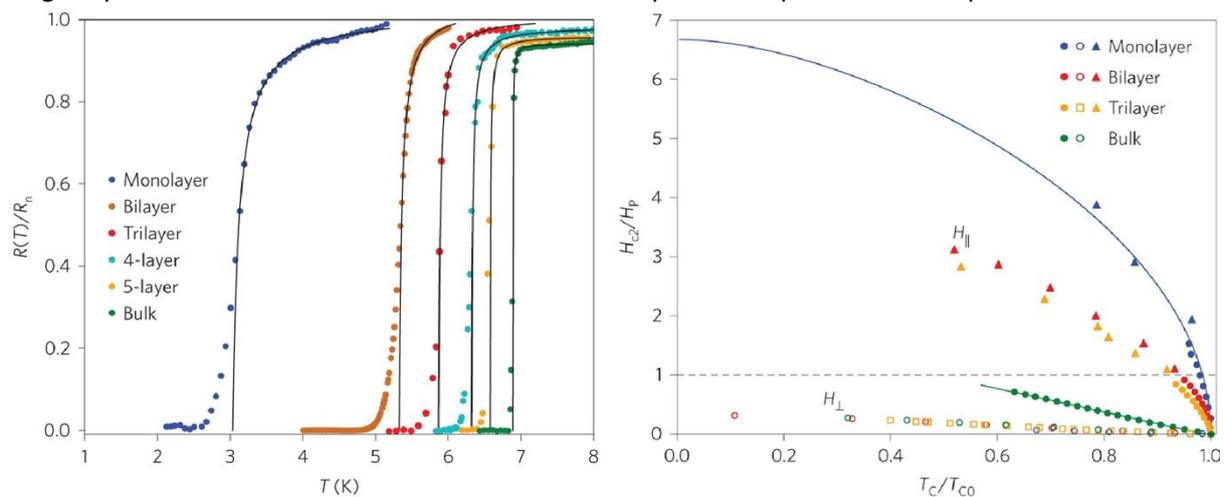

**Figure 2**. Temperature dependence of the resistance (left) and upper critical magnetic fields (right) of NbSe2 with respect to the number of layers. Reprinted by permission from [3].

## Concluding Remarks

Ultrathin layered superconductors are a highly topical field in condensed matter research and the progress that has been made recently is considerable. Apart from proving that robust superconductivity can indeed exist in 2D and verifying concepts like the Berezinskii-Kosterlitz-Thouless transition and quantum Griffiths singularity, researchers have found ways to induce (topological) insulator to superconductor transitions, unravel Ising pairing, and shed light on the coexistence of CDW order and superconductivity, to name just a few mile stones [1, 2]. It is to be expected that this research pace will even further increase with prospects of finding topologically nontrivial superconductivity, the observation and manipulation of Majorana zero modes, and possibly new high-$T_c$ superconductors, to name just a few examples, along this highly fascinating road of research.

**Acknowledgements**

C.H. acknowledges support from the Austrian Science Fund (FWF) Project No. P 32144-N36.

# Topological Insulators

## Adriana I. Figueroa[1] and Bernard Plaçais[2]

[1] Catalan Institute of Nanoscience and Nanotechnology (ICN2), CSIC and BIST, Bellaterra, Barcelona, Spain

[2] Laboratoire de Physique de l'Ecole normale supérieure, ENS, Université PSL, CNRS, Sorbonne Université, Université de Paris, Paris, France.

## Status

First introduced in 2005 by Kane and Mele, topological insulators (TI) constitute a new topological phase driven by spin-orbit coupling (SOC) and symmetries, yielding a bulk insulator with spin-polarised "topologically protected" conducting boundary states (see review **[1]** and references therein). 2D-TIs exhibit the quantum spin Hall (QSH) effect, i.e., a pair of counter-propagating spin-polarised edge states, responsible for quantized conductance at zero magnetic field, as observed in HgTe quantum wells (QW) **[1]** and monolayer WTe$_2$ **[2]**. 3D-TIs sustain massless topological surface states (TSS) with a peculiar spin-momentum locking. The TI family includes a broad class of van der Waals (vdW) materials such as 3D-tetradymites [(Bi,Sb)$_2$(Se,Te)$_3$]. Angle-resolved photoemission spectroscopy (ARPES) **[1]**, as well as magneto-transport and magneto-optics, have provided invaluable insights into boundary states. Current challenges reside in exploiting their superior transport properties for emerging topological electronics and spintronics, which is the focus of this roadmap section.

Zero-field QSH[EB1] is a promising platform for spin-to-charge conversion and superconducting topological junctions. TIs in proximity with superconductors exhibit electronic states that support Majorana fermions, a phenomenon that opens new avenues to achieving topological quantum computation **[3]**. However, conductance quantization (Fig. 1a) suffers from limited spatial robustness, suggesting that topological protection via time-reversal symmetry (TRS) remains fragile. TSS transport in 3D-TIs is more robust, especially in buried topological heterojunctions. Volkov and Pankratov in 1985 anticipated TSS and predicted the existence of massive surface states in smooth heterojunctions (VPs), both subject to a relativistic field-effect limiting Dirac screening **[4]**.

Breaking TRS in 2D and 3D TIs by introducing magnetic order results in the observation of unique quantum phenomena. Magnetic doping, as in Cr- and V-doped (Bi,Sb)$_2$Te$_3$, has led to the advent of magnetic topological insulators (MTIs) that manifest robust QAH effects **[5,6]** (Fig. 1b), topological magnetoelectric (TME) effects **[7]**, and axion electrodynamics. Tailoring and engineering of MTIs and their heterostructures have been also successful in this respect. Exploitation of the TSS, spin texture and strong SOC in devices incorporating TIs has proven their potential for spintronic applications.

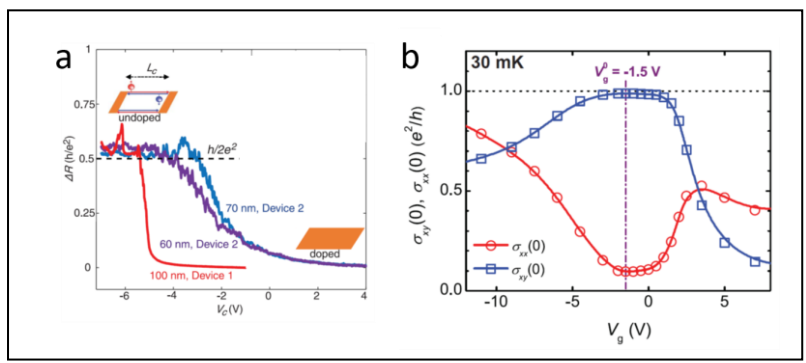

**Figure 1.** a) Quantum spin Hall (QSH) quantization in in the insulating bandgap of single layer WTe$_2$ devices (reproduced from [2]). b) Zero magnetic field quantum anomalous Hall (QAH) quantization in the insulating bandgap of Cr-doped (Bi,Sb)$_2$Te$_3$ (reproduced from [5]).

**Current and Future Challenges**

The relative fragility of QSH quantization in TIs remains unsolved. Today, improved nanofabrication methods comprise cleaner etching techniques that enable the realization of smooth edges where robust QSH transport and superconducting topological junctions in HgTe-QWs at the micrometer scale are observed **[8]**. Despite the advances this achievement entails, it is not yet comparable to the high-precision QAH quantization reported in 100 μm-long Cr-doped (Bi,Sb)$_2$Te$_3$ Hall bars **[9]**. Even if further progress is expected on the technological side, the marked contrast between QSH and QAH quantization raises basic physics questions on the absolute robustness of TRS protection. Possible routes may rely on decoherence and coupling to the environment bath which control the irreversible decay toward thermodynamic equilibrium. Engineering this environment is a key challenge that requires a better knowledge of bulk electronic but also bosonic excitations, such as phonons, plasmons, polaritons, magnons, and the still elusive axion electromagnetic modes.

Similarly, new routes and concepts are required in order to improve the properties of MTIs. The first intrinsic MTI, the layered antiferromagnet MnBi$_2$Te$_4$, was recently discovered, where both QAH **[6]** and axion **[7]** insulator states occur depending on the parity of the number of layers (Fig. 2). Notwithstanding, the low temperatures (< 2 K) at which these and related quantum phenomena are observed in this material, as well as in magnetically-doped TIs, makes it challenging to incorporate them in devices with practical applications. Reducing disorder in the exchange gap is believed to be widen temperature range in which quantum phenomena manifest in these systems. Alternative approaches take advantage of/make use of proximity effects in magnetic heterostructures incorporating TIs to pursue QAH effects, as well as to use chiral edge modes to electrically control the magnetization, or vice versa, exploiting spintronic effects such as spin-orbit torques.

Both in the case of MTIs and non-magnetic TIs, along with their heterostructures, further theoretical studies and experimental investigations are needed to take up the challenge and fulfil the promises of topological electronics and spintronics. The versatility of vdW-TIs offers a unique platform for topological science that is still at its infancy.

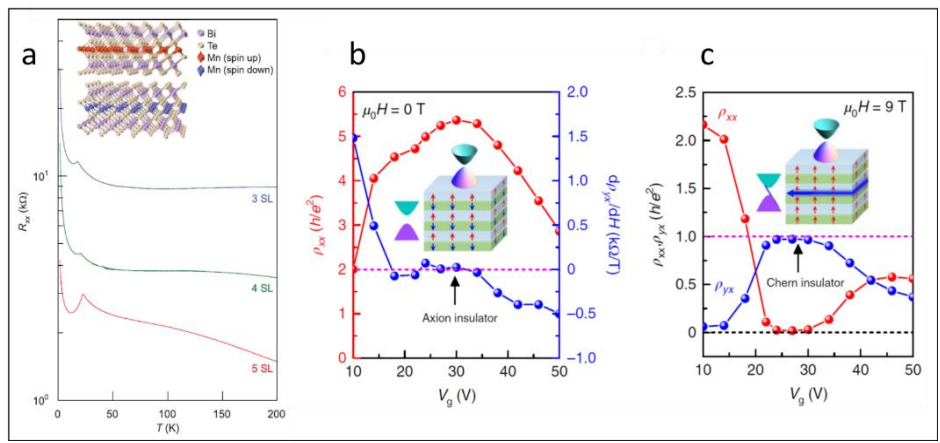

**Figure 2.** a) Temperature dependent sample resistance of few-layer MnBi$_2$Te$_4$. The antiferromagnetic transition manifests as a resistance peak in the three-, four-, and five-layer samples. Inset shows the layered crystal structure of MnBi$_2$Te$_4$ in the antiferromagnetic state. The spins of Mn$^{2+}$ ions order ferromagnetically within a layer, whereas neighboring layers couple antiferromagnetically with an out-of-plane magnetocrystalline anisotropy. Gate-dependent resistance of a six-layer MnBi$_2$Te$_4$ that exhibits b) axion and c) Chern insulator (QAH) states. Insets in (b) and (c) illustrate of the magnetic order and electronic structure of such states. Panel (a) adapted from Ref. [6] and panel (b) and (c) reproduced from Ref. [7].

**Advances in Science and Technology to Meet Challenges**

Topological insulator physics and technology has not reached the readiness level to envision the building up of a genuine "topotronics" foundry. If narrow-bandgap semiconducting TIs (CdHgTe, PbSnTe) are produced at the wafer scale with high mobility, it is not yet the case of medium bandgap vdW-TIs. In the near future, the latter will largely benefit from the intense efforts worldwide to realize large scale, high-quality films and heterostructures with controlled chemical doping. On the device side, a better understanding of the specific field effect in TIs is needed because of the coexistence of trivial and topological carriers. It would deserve a new compendium on the physics of topological semiconducting devices. It is also very desirable to build a comprehensive database on TI-material excitations, in complement to the recent catalogue of high-quality topological crystals **[10]**, as they play a crucial role in limiting finite-bias properties of topological devices. As a matter of fact, TI materials, which aim at being sensitive sensors for thermal, magnetics, far-infrared and microwave excitations, and building blocks of topological quantum computation, are primarily sensitive to their own thermal fluctuations bath. There are possibly more crystalline structures to be synthetized from chemistry among the 3307 TI-crystals identified in Ref. **[10]**, possibly including cleverly engineered metamaterials. On the physics side, the role of interactions in triggering collective instabilities is worth investigating. A theoretical account of actual finite confinement lengths calls for a phenomenological approach of coexisting trivial and topological carriers in the presence of band bending.

On the experimental side, high quality growth and fabrication of TIs and their heterostructures is currently achieved by molecular beam epitaxy (MBE), which continues to be the best technique to produce, for instance, layered materials with superior properties than their bulk crystal counterpart. Optimized MBE grown MTIs, such as $MnBi_2Te_4$ and related compounds, should reduce their disorder and further increase the energy scale of the desired QAH and TME effects. More exotic quantum states may be implemented with heterostructure engineering and appropriate materials combinations. Efficient control of the TME and other spin-related phenomena is one of the main goals in spintronics and would constitute a further step towards industrial applications of TIs.

**Concluding Remarks**

*[Include brief concluding remarks. This should not be longer than a short paragraph. (150 words max)]*

The last 15 years have been intense and very active in research and development of TI materials. Important progress has been done, but further improvements in the quantization, energy and temperature scales are required in order to achieve industrial and technological advances. The gap between fundamental research and practical applications remains open and further strategies are needed to close it. More in-depth theoretical and experimental research is needed to clarify the nature and characteristics of strong SOC effects, in order to guide experiments towards more efficient and practical implementations of spin-related mechanisms.

**Acknowledgements**

*[Please include any acknowledgements and funding information as appropriate.]*
Authors express their deep gratitude to the medical care personnel during the Covid-19 pandemic. B.P. wishes to thank E. Bocquillon for fruitful discussions and suggestions. A. I. F. and B. P. acknowledge funding from the European Union's Horizon 2020 research and innovation programme under respectively the Marie Skłodowska-Curie grant agreement No. 796925, and the grant agreement No. 785219 Graphene Core 2.

## Topological semimetals

QuanSheng Wu and Oleg V. Yazyev

Institute of Physics, Ecole Polytechnique Fédérale de Lausanne (EPFL), Lausanne CH-1015, Switzerland

**Status**

*[This section provides a brief history and status, why the field is still important, what will be gained with further advances. (350 words max)]*

The field of topological materials has seen explosive development during the past decade. While initial efforts focused on materials hosting insulating topological phases ($Z_2$ topological insulators, followed by topological crystalline insulators), it was soon realized that gapless topological phases also present rich and exciting physics [1]. The first relevant topological phase is the Dirac semimetal characterized by the point band degeneracies connecting linearly dispersing bands at the Fermi level. In these materials, which can be viewed as three-dimensional analogues of graphene, the four-fold band degeneracies are protected by the crystalline symmetries. Two notable examples of Dirac semimetals are $Na_3Bi$ [2] and $Cd_3As_2$ [3], both experimentally confirmed in 2014 following earlier theoretical predictions. Breaking inversion or time-reversal symmetries results in two-fold band degeneracies that are chiral, i.e. characterized by non-zero chiral charges (Fig. 1a). Materials in which such point degeneracies are isolated at the Fermi level are referred to as the Weyl semimetals [4]. First members of this broad class of topological semimetals have been discovered in the TaAs family of materials in 2015 (Fig. 2a) [5,6]. The notion of type-II Weyl semimetals was then introduced to account for the case of strongly tilted conical band dispersion with band degeneracies representing the touching points between electron and hole Fermi surface pockets (Fig. 1b) [7]. Examples of type-II Weyl semimetals are the initially proposed ditellurides $WTe_2$ [7] and $MoTe_2$, confirmed robust Weyl semimetals in diphosphides $WP_2$ and $MoP_2$ [8]. Band degeneracies can further extend along a line or a loop (Fig. 1c) leading to another distinct topological phase – the nodal line semimetal – with $PbTaSe_2$ and materials from the ZrSiS family being typical representatives. The classification of topological semimetals does not stop here now including scenarios with even more complex forms of band degeneracies, such as nodal chains, six- and eight-fold degeneracies, and novel classification schemes such as the non-Abelian band topology [9]. Likewise, the number of theoretically proposed and experimentally confirmed materials hosting these phases continues growing at a steady pace. The progress in the emerging field of topological semimetals is driven by two forces – fundamental physics and prospective technological applications. As far the former is concerned, the paradigm of realizing new types of fermions as quasiparticle excitations in topological condensed matter systems establishes an interesting connection with high-energy particle physics. On the other hand, novel properties of topological semimetals are actively investigated for potential applications ranging from catalysis to topological quantum computing.

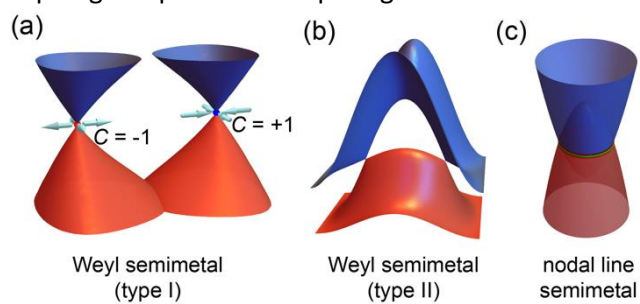

**Figure 1.** Schematic representations of the idealized band structures of (a) type-I and (b) type-II Weyl semimetals, (c) nodal line semimetal. The opposite chiral charges of the two Weyl point degeneracies in (a) are indicated. The nodal line band degeneracy in (c) is shown by the green line.

**Current and Future Challenges**

*[This section discusses the big research issues and challenges. (350 words max)]*

The field of topological semimetals can be regarded as nascent and shows very rapid evolution. Despite the frequently emerging new ideas, the outstanding challenges in the field have been posed since its very beginning. We can identify the following open problems:

(1) Angle-resolved photoemission spectroscopy (ARPES) remains the primary experimental tool for confirming the presence of a topological semimetal phase in a given "candidate" material, while the predictions would typically rest on materials chemistry intuition, symmetry considerations and the results of first-principles calculations. ARPES provides direct access to topological band degeneracies in bulk band structures as well as to the surfaces states resulting from the corresponding bulk topological phases (Fig. 2c,d), e.g. Fermi arcs and drumhead states in the Weyl and nodal line semimetals, respectively. However, observing the anticipated unusual transport signatures of the topological phases, e.g. the chiral anomaly, gravitational anomaly, the planar Hall effect, etc., appears to be more challenging. Despites numerous reports in the literature, there is a growing concern regarding possible alternative explanations of these measurements, or even their reproducibility.

(2) It would not be an exaggeration to say that theoretical predictions have been driving the progress of the field of topological semimetals. Many recent high-throughput computational surveys claim that a substantial fraction of all know crystalline materials host some kind of topological phase. This is not surprising given the fact that breaking the inversion or time-reversal symmetries, or introducing non-symmorphic symmetries, is sufficient for creating topological band degeneracies. The real challenge, however, consists in finding materials having the minimal possible number of such degeneracies isolated at the Fermi level, in correspondence to the ideal models of topological semimetals. Currently investigated materials, such as the TaAs-family Weyl semimetals, are still far from the ideal models showing very complex arrangement of topological band degeneracies (Fig. 2b). A close dialogue between theory and experimental researchers is necessary for overcoming this challenge.

(3) Most currently investigated topological semimetals are weakly correlated materials. It is anticipated, however, that the combination of topology and strong electron-electron interactions will open vast opportunities for discovering novel physical phenomena. Finding a convenient platform material that realizes such interplay between strong correlations and topology as well as focusing both theoretical and experimental efforts on a particular important phenomenon is a challenge that needs to be addressed in the coming years.

(4) Currently, the progress in the emerging field of topological semimetals is largely driven by the exploration of novel physics rather than by prospective technological applications of these materials. Identifying an application domain where topological semimetals will play the role of enabling materials, finding a material that could be considered the "silicon" of topological semimetals and developing device manufacturing techniques defines another challenge for this emerging field.

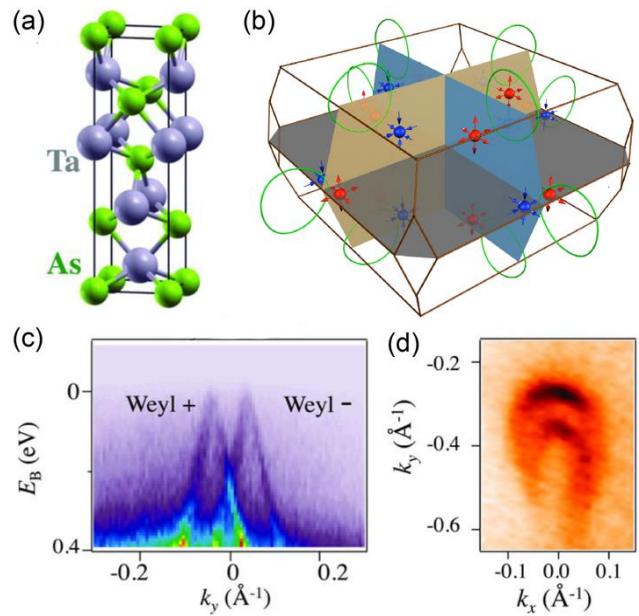

**Figure 2.** (a) Body-centered tetragonal crystal structure of TaAs (space group *I*4₁*md*, No. 109). (b) Positions of the 24 Weyl points of two distinct types in the Brillouin zone of TaAs predicted by first-principles calculations [5]. Experimental observation of (c) the Weyl nodes add (d) the surface Fermi arcs in TaAs using ARPES [6]. Panels (a),(c),(d) adapted with permission from [6]. Copyright (2015) AAAS. Panel (b) reproduced with permission from [5]. Copyright (2015) APS.

## Advances in Science and Technology to Meet Challenges

*[This section discusses the advances in science and technology needed to address the challenges. (350 words max)]*

The field of topological semimetals is still very young, but it quickly advances towards reaching maturity. It is currently flooded with a large number of proposed novel physical phenomena, different kinds of topological phases as well as candidate materials hosting them. Natural evolution of this domain will soon lead to identifying the most interesting phenomena and materials, thus making scientific progress in this field more coherent and focused. One can expect that for selected topological semimetals progress in developing growth methods compatible with common device manufacturing strategies will enable exploring the technological potential of these emerging materials. First steps in this direction have already been reported [10].

## Concluding Remarks

*[Include brief concluding remarks. This should not be longer than a short paragraph. (150 words max)]*

The field of topological semimetals and the broader domain of topological states of matter are currently among the most rapidly expanding areas of physics. One can expect that very soon topological materials with have their own chapter in introductory condensed matter physics textbooks, at the same level as semiconductors, superconductors or magnetic materials. The grand challenge of the field, however, consists in realizing first paradigm shifting technological applications of the newly discovered physical phenomena and materials. Success in achieving this goal largely depends not only on collaborative research in theoretical and experimental physics, but also on extensive involvement of researchers in chemistry, materials science and engineering disciplines. These adjacent fields are only starting to realize the perspectives of topological materials.

## Acknowledgements

*[Please include any acknowledgements and funding information as appropriate.]*

We would like to thank Philip Moll for discussions and NCCR Marvel for support.

**References**

*[(Separate from the two-page limit) Limit of 10 References. Please provide the full author list, and article title, for each reference to maintain style consistency in the combined roadmap article. Style should be consistent with all other contributions- use IEEE style]*

## Quantum Materials for Topological Devices based on Majorana modes

Erik P. A. M. Bakkers (Eindhoven University of Technology) and Jesper Nygård (Niels Bohr Institute, University of Copenhagen)

**Status**

Majorana modes are quasi-particle states that can exist at interfaces in certain classes of materials. The states are inherently non-local and the emergence of a split pair of Majorana states is a remarkable manifestation of topology, an important new paradigm in condensed matter physics with direct links to materials science [1]. The observation and manipulation of these exotic quasi-particles are of great scientific interest; for instance they do not obey ordinary (fermionic) exchange statistics. Topological states entail schemes for realizing quantum information processing based on qubits that are in principle protected from decoherence. This advantage over other solid state qubits has not yet been demonstrated, but electrical transport experiments have shown signatures of the essential Majorana modes that potentially allows for encoding of the topologically protected qubit states.

We focus here on electronic semiconductor devices while noting that other materials have been studied e.g. by STM [1]. The Majorana modes emerge as degenerate, zero-energy states and they are predicted to exist at the ends of one-dimensional conductors in a topological superconducting state [2] (Fig. 1a). Until date most experiments have been based on III-V semiconductor nanowires (InAs, InSb) grown by vapour-liquid-solid growth (VLS) [3]. In these semiconductors spin-orbit interactions are strong and they can readily by interfaced with superconducting materials – key ingredients needed for realizing the topological state. The materials have been refined by development of hybrid materials where superconducting layers are grown directly on semiconductors in-situ. Some elemental superconductors can be matched epitaxially to facets of semiconductor nanowires (Fig. 1b), facilitating transfer of the superconducting correlations uniformly into the semiconductor nanowire.

The first promising results were observed in two-terminal electrical measurements on nanowires, configured to allow for tunnelling spectroscopy of the electronic states in a semiconductor segment contacted by a superconductor [4]. Here, the desired Majorana zero-energy modes (MZMs) were reported to appear as zero-bias peaks (ZBPs), i.e., distinct signals in electrical transport when electrons could tunnel into a Majorana state. A typical device is shown in Fig. 1c. Such experiments merely indicate the existence of the MZM [5]. While other concepts such as quantum dot and Josephson effect spectroscopies have also been implemented [3,5], none of these experiments allow to test the most profound consequence of topology, the non-Abelian exchange statistics. This would require tunable junctions between multiple one-dimensional nanowires. This need for more complex devices has stimulated significant efforts on developing new nanostructured hybrid materials.

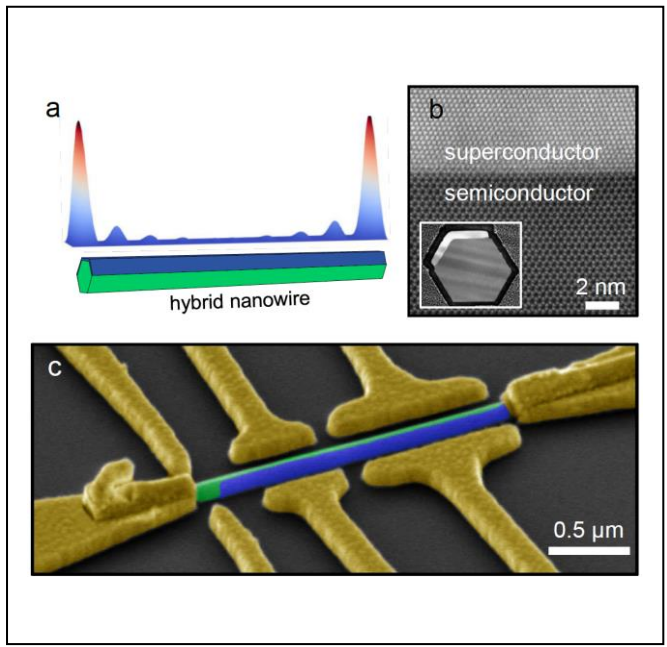

**Figure 1.** (a) Simulation of Majorana modes at the ends of a wire segment in a topological superconducting state, induced by the proximity between a semiconducting nanowire core (green) and a superconducting shell (blue). The topological protection of the Majorana mode is linked to the spatial separation of its two components. (b) Transmission electron micrograph of the epitaxial interface between the two materials and cross sectional view of a hybrid nanowire (inset) with a semiconductor core (dark) and superconducting shell (light). (c) Typical device geometry for tunneling spectroscopy of Majorana modes, with source/drain electrodes contacting the wire and several electrostatic gate electrodes. (Panel (a,c) adapted from M.-T. Deng et al., Phys. Rev. B 98, 085125 (2018)), (b) T. Kanne, L. Zheng, E. Olsson et al. (unpublished).

## Current and Future Challenges

ZBP signals have been reported, but despite much effort there is no conclusive evidence for MZM. A braiding experiment should reveal the non-Abelian statistical properties, and at the same time could embody a first Majorana qubit. This is a challenging experiment, and currently out-of-reach due to the required device quality and complexity [8] as will be discussed below. A more accessible experiment, and a stepping-stone towards braiding, is to demonstrate correlations between two ZBP states at each side of a topological superconductor (Fig. 1a). Correlated states showing all properties as expected from MZM will provide strong indications for the existence of Majorana modes.

The challenge to be addressed for these correlation experiments is the quality of the devices. It has been argued that the ZBPs observed so far could have been induced by disorder [5]. Disorder can be present in the nanowire channel or at the semiconductor/superconductor interface. Disorder in the semiconductor channel is probably induced by the nearby surface and could be reduced by moving surface states further away by using high quality shells, or by using 2DEGs, which are pinched off to form a 1D channel. Challenge of the materials used in these studies is their large lattice constant and the inherent difficulty in finding suitable lattice-matched materials that could serve as a substrate or shell. The quality of the semiconductor/superconductor interface has been greatly improved by epitaxial growth (Fig. 1b) [6]. A new problem arising from epitaxial growth is that the semiconductor channel is 'metallized' by the superconductor due to the much higher density of states. A solution could be to employ a tunnel junction to tune the semiconductor/superconductor coupling. Another approach to mitigate the effect of disorder is by increasing energy scales of the topological system, which would require new materials.

Although the clearest ZBP signals have so far been obtained with bottom-up grown VLS wires, this system is not scalable towards larger circuits with complex device geometries (Fig. 2a). There is a great challenge in developing a scalable material platform, preferably integrated in Si-technology, which can also be coupled to other types of qubits (for universal quantum computing).

**Advances in Science and Technology to Meet Challenges**
Different routes have been identified towards scalable structures. Most prominent are based on two-dimensional electron gasses [7] and selected area grown (SAG) structures [8]. These planar platforms are suited for more complex device geometries, since structures can be defined by top-down approaches using lithography, see Fig. 2b. Integration in Si-technology could possibly be realized by a SAG approach, which recently has proven to be successful in growing high quality structures for optical applications. However, it is more challenging with these device geometries to control the chemical potential with bottom-gates.

As seen above several of the key challenges in observing and utilizing MZM are related to intrinsic material properties. An important direction in improving device quality is to fabricate semiconductor/superconductor interfaces in-situ under ultra-high vacuum and at low temperatures. This results in cleaner systems with sharp interfaces [6]. In-situ shadow structures is a powerful method to eliminate the need for etching or lift-off, as has been demonstrated for VLS based nanowire geometries and will be expanded towards SAG-based technologies [8,9]. In addition, methods are being developed for complete in-situ device fabrication, including normal contacts, magnetic materials, and dielectrics to realize pristine devices.

A third important development is in exploring new materials which increase the energy scale. The in-situ growth approach allows examining a number of promising superconductors with higher critical fields [9]. In addition, topological insulators (TI), such as $Bi_2Te_3$ and SnTe, with relatively large bandgaps are of great interest [1,10]. Despite much effort, the challenge is still to find TIs which have an insulating bulk.

Braiding operations on MZMs would be suitable for encoding single qubit gates in quantum computing applications. Initial braiding proposals were based on physically moving Majorana modes in nanowire networks, but recent theoretical advances have led to promising schemes where manipulations can be obtained by adiabatic control of the interactions between MZMs or projective measurements [11]. Here both read-out and control can be performed by the same units, for instance quantum dots coupled to MZMs (see Fig. 2a). We note that only single qubit gates can potentially be realized with MZMs. Universal gates require that these are supplemented with other types of qubit operations. Thus topologically protected universal quantum computing is not only a matter of scaling up MZM systems.

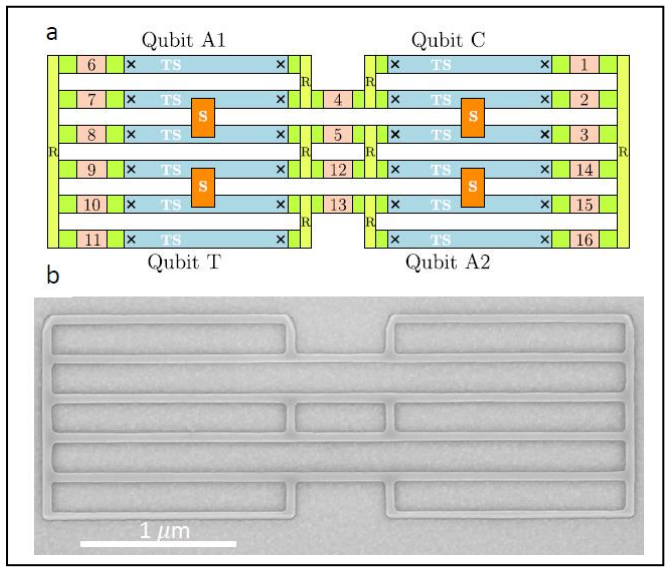

**Figure 2.** (a) Envisioned four qubit structure based on 'Majorana box qubits', comprising long sections of topological superconductor (TS, blue), which are shunted by superconducting bridges (S,,orange), few-channel semiconducting reference arms (R, light green ), normal metals (yellow), and gate-controlled quantum dots (light red) [11]. (b) Scanning Electron Microscopy image of single crystalline InSb structure realized by selective-area growth on an insulating InP substrate based on the design shown in (a) [8].

## Concluding Remarks

We have reviewed concepts and materials aspects behind the advances in Majorana-based devices. During the past decade academic groups and corporate labs have jointly put significant efforts into this research. The understanding and refinement of the necessary systems has progressed and there is mounting evidence for the existence of MZMs in solid state devices. Yet, a Majorana-based qubit has not been demonstrated but it is clear that progress on topological quantum computing must go hand-in-hand with advances in materials science. We anticipate that the key topological properties of MZMs will be demonstrated the next few years. However, implementing these in complex circuits is a daunting challenge requiring heterostructures engineered with a precision at the atomic scale. Yet, these efforts have already opened up a fascinating area of science.

## Acknowledgements

EB has received funding from the European Research Council (ERC) under the European Union's Horizon 2020 research and innovation programme (TOCINA, grant agreement No. [834290]). JN acknowledges funding from the Danish National Research Foundation Center for Quantum Devices and the Carlsberg Foundation.

[1]Institut de Física d'Altes Energies (IFAE), The Barcelona Institute of Science and Technology (BIST), Bellaterra (Barcelona) 08193, Spain

[2]Qilimanjaro Quantum Tech, Barcelona, Spain

[3]Univ. Grenoble Alpes and CEA, IRIG/PHELIQS, F-38000 Grenoble, France

[1]Institut de Física d'Altes Energies (IFAE), The Barcelona Institute of Science and Technology (BIST), Bellaterra (Barcelona) 08193, Spain

[2]Qilimanjaro Quantum Tech, Barcelona, Spain

[3]Univ. Grenoble Alpes and CEA, IRIG/PHELIQS, F-38000 Grenoble, France

# Superconductor and Semiconductor Qubits

Pol Forn-Díaz[1,2], Silvano De Franceschi[3]

May 8, 2020

## 1 Introduction

Quantum computation is possibly the quantum technology with the highest transformative potential. In order to implement a quantum processor in a realistic physical platform, several requirements need to be satisfied: long qubit coherence, qubit state initialization capability, a universal set of gates with high fidelity, an efficient qubit readout, and finally, and most crucially, a scalable physical architecture.

Achieving long coherence times requires a profound understanding of the physics of the environment surrounding the qubits, particularly in the solid-state implementations discussed in this section. Eventually, qubit coherence sets a limit on gate fidelity. Incidentally, a minimum gate fidelity is required before quantum error correcting codes may be implemented to realize universal quantum computation.

Identifying a truly scalable physical platform is the most important challenge faced in the field. Implementations of just a few qubits may not show limitations appearing at larger system size. Scalability requires a truly technological boost to advance any physical platforms to become quantum processors. In recent years, the jump to this different way of operation has led to large industries take over the fundamental research developed at academic centers.

In this section, we focus on two types of solid-state architectures: superconductor- and semiconductor-based qubits. These two physical platforms are currently the most promising in realising quantum computing protocols with a solid state system. Superconducting qubits have shown robustness in scaling up the system size to the current world record 53-qubit device developed by Google [1]. Semiconductor-based qubits have also shown progress in recent years, with their true potential being the accessibility to integrate to the classical transistor technology, enabling the scalability of this platform. In addition, semiconductors can be interfaced with light to enable quantum communication to develop quantum networks.

## 2 Current and future challenges

### 2.1 Superconducting qubits

Superconducting materials have since their appearance represented an exciting technological opportunity. Their limited widespread use lies in the difficulty to combine the low temperatures required to operate them in real-life applications. Nevertheless, several important technologies entirely rely on superconductors, such as magnets in magnetic resonance imaging or particle accelerators. Other emerging applications may soon become a reality, such as marine cables to transfer energy generated in offshore wind turbines or faster and less energy-consuming digital supercomputers.

The non-intuitive properties of superconductors have also attracted much interest to scientists to study their fundamental quantum properties. Already in the 80s, Nobel Laureate Anthony Leggett played with the idea to control the macroscopic quantum state of a superconducting circuit exhibiting flux quantization. Late in the 90s, experiments were already at the level of exploring quantum coherent effects in devices where charge or flux quantization dominated the dynamics. Those experiments represented the birth of superconducting qubits.

The early qubits showed poor coherence properties due to their enhanced sensitivity to charge and magnetic flux fluctuations. It was not until 2004 when microwave engineering was properly introduced in superconducting quantum circuits to yield circuit quantum electrodynamics, circuit QED in short, pioneered by the Yale team [2]. This revolutionary architecture permitted separating material losses from those originating in the control and readout circuitry, as the latter were now filtered out by a resonant circuit protecting the qubit. In addition, circuit QED offered a clear path to scaling up the system size, as confirmed in successively larger devices all the way to the 53-qubit device engineered by the Google team [1]. This architecture is already being adopted by other qubit types, such as spin qubits [3].

The leading superconducting qubit types (see Fig. 1) include the transmon qubit, consisting of a single junction shunted by a capacitor to maximize charge insensitivity [4], the flux qubit, a qubit sensitive to magnetic fields, and the fluxonium qubit, a single junction shunted by a superinductance [5], which is a precursor of topologically protected qubits, such as the $0 - \pi$ qubit [6].

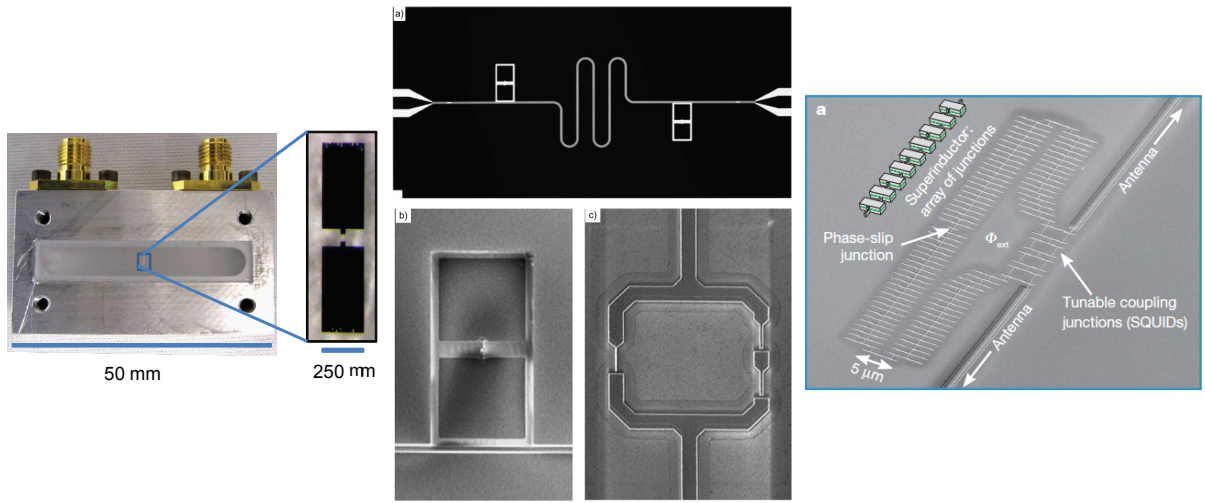

Figure 1: Superconducting qubits. Left: Transmon qubit in its 3D version in an aluminum cavity. Reproduced with permission from [7]. Center: Flux qubit with capacitive shunt developed at MIT, shown with its measurement resonator. Reproduced with permission from [8]. Right: Fluxonium qubit with its Josephson junction array acting as superinductor. Reproduced with permission from [5].

The quality of superconducting qubits has improved by five orders of magnitude since their inception, from nanosecond coherence times to over 100 $\mu s$ [9]. At present, single-qubit gates reach fidelities better than 99.9%, while two-qubit gates are in the 99.8% level [1]. Despite the remarkable progress, the field faces important challenges, mostly related to material losses originated in the surfaces and interfaces of the device that limit coherence times and introduce instabilities from small but noticeable qubit parameter fluctuations. An order of magnitude improvement is required before quantum error-correcting techniques are implementable and the system can be scaled to run as a universal quantum processor. We list here the most relevant challenges:

- Qubit fabrication techniques. Since their inception, superconducting qubits have been defined with aluminum circuits. The key circuit element is the Josephson tunnel junction, usually consisting of an aluminum oxide tunnel barrier between two aluminum electrodes. The Dolan bridge technique with evaporated aluminum is the most widely used procedure, although alternative methods are being put forward, such as overlap junctions [10]. Remarkably, little has been modified in the fabrication of Josephson junctions since the early days. The amorphous nature of the junction electrodes introduces a spread in critical current values, usually in the range of 5%-10%. Reducing this number is a challenge to overcome when very large-scale systems are built and a low parameter yield is not tolerated.

- Superconducting materials. Resonators are often fabricated en-par with the same qubit material, aluminum, but often can be produced from superconductors with higher critical temperature and better intrinsic quality factors at the single-photon level such as niobium (Nb), niobium titanium nitride (NbTiN), or titanium nitride (TiN). The robustness of resonators makes them an ideal test bed for material losses that also affect qubits [11]. The use of resonators has singled out sapphire and intrinsic silicon as ideal substrates for qubit circuits. Finding superconducting materials better than aluminum is a direct path to improve qubit quality. Materials with lower microwave losses may reduce the surface contribution to decoherence. Recently [12], a team at Princeton has demonstrated qubits with record coherence times $T_1 = 362$ $\mu s$ and $T_{2,Ramsey} = 105$ $\mu s$ using tantalum and in-depth surface cleaning procedures. The insulating oxide in tantalum reduces microwave losses compared to other superconductors such as niobium. Even better materials may be soon discovered.

- Two-level defects. All experimental evidence points to the majority of noise felt by superconducting qubits originated in the surfaces and interfaces [9]. The interaction with surface losses was decreased by diluting the electric fields in large capacitors shunting the qubits [4, 8], leading to the most significant improvement in qubit coherence in the last decade. Despite the steady increase in qubit lifetimes, the coherence times have not been enhanced significantly since 2011 when three-dimensional cavities were introduced [7]. Another important limitation attributed to two-level defects is the observed fluctuation of the intrinsic qubit parameters, such as the qubit frequency and the qubit coherence times [13], requiring intensive calibration methods [1]. All these deficiencies are the net effect of microscopic physics with its own dynamics in close contact to the qubit states [14]. This environment turns out to be of a rather complex nature and it responds as a $1/f$ spectrum of both electric and magnetic flux fluctuations [15]. The responsible of these fluctuations remains an area of active research, but evidence seems to point to weakly physisorbed atomic hydrogen together with free radicals originated from water-based solvents during the device production. Initial methods to mitigate their presence, such as annealing in vacuum, have been put in place [16]. Therefore, a significant amount of work is still required to passivate in some way the effect of the substrate noise to further enhance the qubit coherence and stabilise its parameters.

## 2.2 Semiconductor qubits

Semiconductors form the core materials for microelectronics and optoelectronics. The associated technologies have reached a high-level of maturity enabling large-scale manufacturing. In semiconductors, the position and motion of charge carriers (i.e. electrons or holes) can be controlled down to the single-charge level using locally applied electric fields. This possibility has enabled the realization of quantum devices in which carriers are confined to form low-dimensional systems with quantized energy levels. In particular, quantum dots are zero-dimensional objects where carriers are confined in all spatial directions resulting in a discrete energy spectrum (typically $0.1 - 10$ meV, depending on material and size). Due to the local Coulomb repulsion, charge occupation in quantum dots is also quantized, provided temperature is lowered below the characteristic charging energy (typically $1 - 10$ meV). Following technical advances in the last couple of decades, it is now routinely possible to obtain quantum dots confining only one electron or one hole.

The spin degree of freedom of a single particle in a magnetic field provides a natural two-level system to encode an elementary bit of quantum information. This idea lies behind the seminal proposal of Loss and DiVincenzo who envisioned a quantum-computing device based on electron spins in quantum-dot arrays [17]. Around twenty years ago, this proposal triggered a world-wide experimental endeavor focusing initially on few-electron quantum dots made in GaAs-based heterostructures [18]. The first spin qubits were reported in 2005. Soon, however, scientists realized that the unavoidable hyperfine interaction with the nuclear spins of the host crystal causes a rapid dephasing of the electron spin state on a time scales of only tens of nanoseconds. This serious problem can be avoided by turning to semiconductor materials with zero nuclear spin. That is the case of group-IV semiconductors (e.g. C, Si, and Ge) whose most abundant isotopes are nuclear-spin free.

Presently, research is mostly focusing on silicon, mainly because of its well-established technology. In addition, silicon offers a range of opportunities for storing and controlling quantum information. Back in 1998, Kane proposed the realization of spin qubits encoded in the nuclear spin of phosphorous dopants properly implanted in a silicon device [19].

The first silicon qubits were reported in 2012 [20, 23]. A few years later, the advent of the first devices made from isotopically purified silicon was a turning point, leading to the development of electron-spin qubits with much longer coherence times. To date, the best silicon qubits have surpassed the threshold for quantum error correction, scoring single-qubit fidelities above 99.9% [24, 25]. The implementation of two-qubit gates is still at an early stage with only a few reported demonstrations [21, 26, 27, 28]. While a two-qubit fidelity of 98 % has been reached so far, ways to go well beyond have been identified [27].

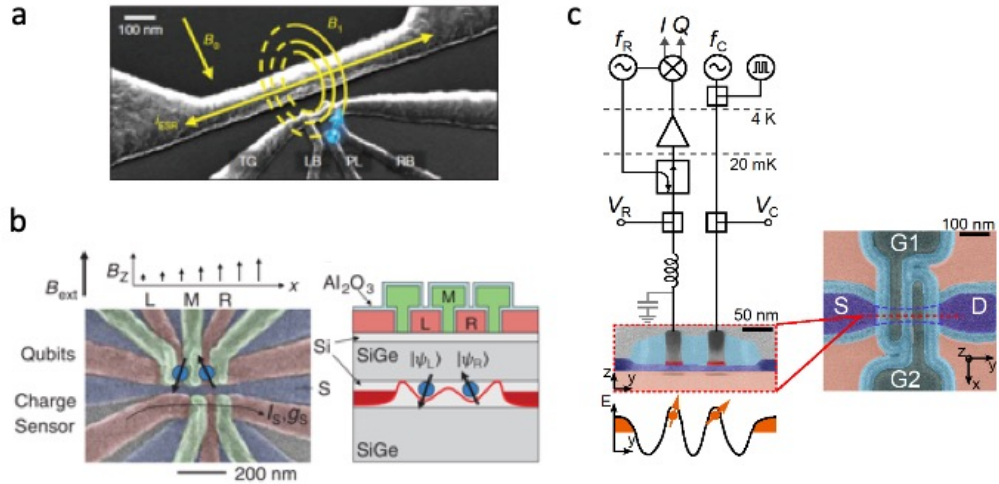

Figure 2: a) Silicon MOS-type device used for the first demonstration of a donor-based qubit. The electron spin bound to a locally implanted phosphorus dopant is magnetically controlled by an ac current through a nearby metal line. Reproduced with permission from [20]. b) Double quantum-dot device used to demonstrate a resonantly driven CNOT gate. Two electrons are electrostatically confined in a buried silicon channel sandwitched between SiGe barrier layers. Their spins are individually controlled by ac gate voltages inducing oscillatory lateral motion in a slanting magnetic field generated by a nearby micromagnet (not shown). Reproduced with permission from [21]. c) Double quantum-dot device made in a 300-mm industry-standard fabrication platform. It was used to demonstrate a compact hole spin qubit with spin-orbit-mediated gate-voltage control and gate-based rf-reflectometry readout. Reproduced from [22].

Following this remarkable progress, silicon-based qubits have become serious contenders for the development of quantum processors. Their potential relies on the possibility of leveraging the large-scale integration capabilities of silicon technology. In this prospect, efforts have begun to focus on the implementation of qubit functionalities in devices made in industry-standard fabrication platforms [22].

Different architectural schemes have been recently proposed to face the scalability challenge [29, 30]. These schemes involve strategic choices based on the current state-of-the-art. Given the multifaceted nature of semiconductor spin qubits, which is far from being thoroughly explored, an evolution of these initial ideas can be expected in compliance with the forthcoming progress. At the level of physical system, control toolbox, and material, a variety of alternative options are currently being benchmarked against each other in order to identify the most promising one with respect to scale-up. We list some of the most relevant ones:

- Electron- vs hole-spin qubits: electron quantum dots exhibit lower-disorder and lower variability. Electron spin qubits have longer coherence times and have shown both one-and two-qubit functionality with the highest fidelities. Hole spin qubits do not suffer from the valley degeneracy problem. Hole spins enable all-electric fast control [22, 31, 32].

- Dopants vs quantum dots: Dopants allow for longer coherence times [33]. Their atomically-precise positioning, either via local implantation or scanning probe lithography is challenging [34], while quantum dots are deterministically defined by gate voltages. Quantum dots show so-far a clear advantage with respect to implementing inter-qubit coupling and creating multi-qubit arrays [35, 36].

- Single-electron vs multi-electron qubits: Encoding a qubit onto multi-electron electron spin states (e.g. singlet and triplet states) is a way to enable electric-field controlled qubit rotations [37]. This, however, has a price in terms of device complexity making scalability more cumbersome with respect to the Loss-DiVincenzo spin-1/2 encoding.

- Magnetic- vs Electric-field control: Electric-dipole spin resonance exploits either intrinsic [38] or synthetic [39] spin orbit coupling. While it allows for faster all-electrical spin manipulation, the spin qubit becomes sensitive to charge noise, which degrades its coherence. Magnetic-field-drive electron spin resonance applies well in the absence of spin-orbit coupling but qubit manipulation times are typically longer.

- Si MOS vs SiGe heterostructures: quantum dot qubits defined in a Si MOS structure are more readily compatible with main-stream CMOS technologies. In addition, the stronger electric fields in MOS structures favor the lifting on valley degeneracy. Quantum dots defined in a buried SiGe/Si/SiGe strained quantum well show a lower level of disorder. Recently, SiGe/Ge/SiGe confining high-mobility holes are also emerging as a promising alternative system combining low disorder [40] and electric-field spin control [32].

# 3  Transverse research developments

The stringent requirements to scale up the size of any quantum system unavoidably leads to novel enabling classical and quantum technologies that may be suited to several physical implementations. In particular, there already exist transverse, enabling classical and quantum technologies developed for superconducting qubits that can be applied to semiconductor qubits. We list some of the most interesting ones here:

1) **Josephson parametric amplifiers for fast high-fidelity spin readout**. The development of both narrow and, especially, wide band near quantum-limited superconducting parametric amplifiers [41] to simultaneously readout multiple qubits is a necessary step in ramping up the complexity of devices. This technology can be applied to any solid-state quantum computing platform and its operation can extend to sub-GHz frequencies, which is interesting for semiconductor qubits [42].

2) **High impedance circuits**. High-impedance resonators based on Nb, NbN, NbTiN, or arrays of superconducting quantum interference devices have been employed for fast single-shot spin readout and for the coherent coupling of distant spin qubits [43].

   In parallel, an old superconducting material known as granular aluminum has been introduced in superconducting qubit technology in recent years [44]. The material is endowed with genuine properties which find applications in quantum technologies with superconducting circuits: high kinetic inductance, low-level losses and high critical magnetic field, on the order of several $T$. The most important is the high kinetic inductance per square, ranging up to several nH/square, which combined with low losses result in a true superinductor with impedances well above the resistance quantum, $R_Q = h/(2e)^2 \approx 6.5$ k$\Omega$. This material is analogous to superinductors obtained by large area Josephson junction arrays used in wide-band parametric amplifiers and fluxonium qubits. Although intrinsically small, the nonlinearity of the material itself can be engineered large enough to be an alternative to Josephson junctions, as already demonstrated in a transmon-like qubit combining aluminum and granular aluminum. This material may play a relevant role in exploring light-matter interactions in the ultrastrong coupling regime [45].

4) **Classical control cryogenic technology**. Superconducting digital circuits have been known for many decades. Based on single flux quantum ideas, classical transistor-based logic can be implemented, providing very fast and very low power electronics. This kind of technology is advancing to become an essential component in any solid-state quantum computing platform, already extensively used by the company D-Wave producing quantum annealers with over 2000 qubits [46].

Conversely, some semiconductor-based developments may be beneficial to superconducting qubits. We list here two examples:

1) **Hybrid superconductor-semiconductor qubit: Gatemon**. Groups in Copenhagen [47] and Delft [48] have developed a superconducting transmon qubit, the so-called gatemon, whose transition frequency can be tuned by a gate voltage. This type of qubit incorporates a Josephson junction consisting of an InAs semiconductor channel with superconducting Al contacts grown in-situ. The Josephson energy, directly linked to the qubit frequency, depends on the carrier density in the semiconductor channel, which is gate dependent. Gatemon coherence times of about 10 $\mu$s have been reported, as well

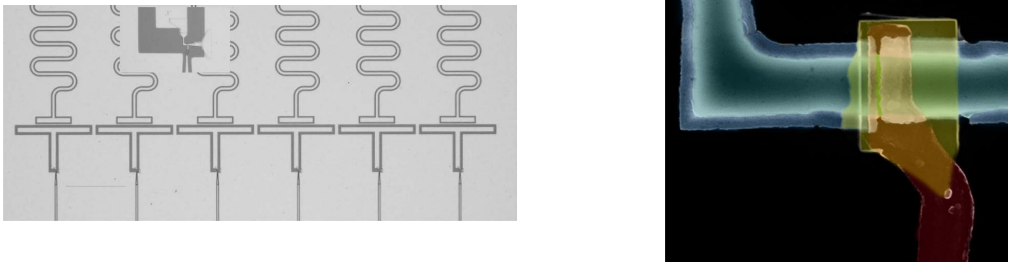

Figure 3: Gatemon qubit. Left: Gatemon array. Right: Closeup of the gatemon junction with an electrode to tune the static potential and change the tunnel barrier electrically. Reproduced with permission from [47]

as single- and two-qubit gate operations [47]. In contrast to flux-tuned transmons, voltage-tunable gatemons may offer a valuable path to large-scale qubit integration.

2) **Cryo-CMOS**. Classical cryogenic electronics based on silicon CMOS technology is a natural choice for semiconductor-based qubits. Since recently, cryo-CMOS is being applied also to superconducting quantum hardware. For instance, the Google team recently developed a first prototype of quantum-control circuit based on 28-nm CMOS technology [49]. The circuit was mounted on the 3-K stage of a dilution refrigerator to perform XY gate operations on a small set of transmon qubits at base temperature.

# 4 Concluding remarks

The most dramatic improvements in superconducting qubit technology to produce working qubits have come from the circuit design surrounding the qubit, and the quality of the materials of the circuit, the substrate, and the interface between them. The challenge of obtaining a stable enough superconducting qubit has to be faced principally from a more profound understanding of the microscopic processes occurring in the neighbourhood of the qubit. Finally, the challenge of further increasing the system size will put even more stringent demands on technological advancements, both classical and quantum mechanical.

Progress toward a semiconductor-based quantum processor will hinge upon finding the best compromise in terms of high one- and two-qubit fidelities, low variability, and scalability of the qubit design. Extending the coherence time and the fidelity of qubit operations will require a tremendous engineering effort at the level of materials (optimization of gate stacks, material interfaces, isotope engineering, dopant incorporation, etc.) and device design (maximal control fields, minimized effect of charge noise, etc.). Increasing the number of qubits will necessarily involve the development of ad-hoc control/readout hardware. As envisioned in the aforementioned proposals [29, 30], elements of this hardware may lie close to the qubits. In this prospect, the development of cryogenic CMOS electronics sharing the same silicon technology is particularly appealing since it would allow for a certain level of quantum-classical co-integration [50].

Eventually, a combination of the technological advances in superconductor- and semiconductor-based qubits may be the path to a hybrid solid state universal quantum computer.

# 5    Acknowledgement

P. F.-D. acknowledges support of a Junior leader fellowship from "la Caixa" Foundation (ID100010434) with code LCF/BQ/PR19/11700009, funding from the Ministry of Economy and Competitiveness, through contracts FIS2017-89860-P and Severo Ochoa SEV-2016-0588, by the QuantERA grant SiUCs by the Agencia Estatal de Investigación with project code PCI2019-111838-2, and by the "la Caixa" Foundation, under MISTI program through agreement LCF/PR/MIT17/11820008. S.D. acknowledges support from the European Union, through the Horizon 2020 research and innovation program (Grant Agreement No. 810504), and from the Agence Nationale de la Recherche, through the CMOSQSPIN project (ANR-17-CE24-0009).

## X – Non-equilibrium phenomena in quantum materials

J. W. McIver, Max Planck Institute for the Structure and Dynamics of Matter, Hamburg, Germany

L. E. F. Foa Torres, FCFM, Universidad de Chile, Chile

**Status**

Much of our technology relies on the non-equilibrium control of solids, from dynamically switching transistors in logic circuits, to reading and writing memory devices, to various energy harvesting applications [1]. This has bred interest in exploring solid-state phenomena that are inherently non-equilibrium, with an eye on creating functionalities that can be manipulated efficiently and at high speeds. Advances harnessing light-matter interaction using ultrafast laser pulses are shifting materials research in this direction, where remarkable quantum effects have been discovered that are already leading to device applications.

A notable early success was the sub-picosecond demagnetization of a ferromagnetic metal [2]. Multiple ultrafast thermal and non-thermal pathways to optically control magnetism have since been discovered [3], and these effects are now being used in next-generation magnetic storage devices [4].

More exotic phenomena can be found in complex solids. In these systems, optical excitation can shift the delicate balance between coexisting or competing ground states, inducing non-equilibrium states with distinct electrical, magnetic and structural properties from their equilibrium counterparts (Fig. 1a) [5-7]. In most cases, these states are transient and decay on the order of femtoseconds to nanoseconds. However, in some cases, persistent metastable phases are obtained that survive indefinitely on experimental timescales [8-10]. $1T\text{-}TaS_2$ is an interesting example, where the insulating ground state and a metallic metastable state can be independently triggered-and-held by using different color laser pulses [10]. This unusual functionality is now being considered as a memory device in future low-temperature electronics.

In recent years, advances in laser technology have enabled the selective excitation of collective modes in quantum materials. Resonantly driven phonon modes, for example, can induce directed lattice distortions that have served as a control knob for ferroelectricity [11-12], magnetism [13] and insulator-to-metal transitions [14]. But perhaps the most striking example is light-induced superconductivity [15], where the optical signatures of superconductivity have been observed in materials well above their equilibrium transition temperatures. This phenomenon has been observed in multiple cuprate and organic superconducting compounds [15-17], suggesting that it may be a common effect associated with strongly correlated superconductivity.

The non-equilibrium dynamics induced by ultrafast optical stimulation can also reveal important information about the ground state of quantum materials, such as the presence of coexisting phases, electron-boson coupling strengths, and the magnitude of correlation-induced energy gaps [18,19]. Pulsed excitation can even initiate oscillations of a phase's order parameter (Fig. 1b), as has been observed in charge density wave systems [18-20], superconductors [21] and excitonic insulators [22]. This phenomenon provides a novel way to map the free energy landscape of solids, and with further research could serve as a reliable probe of order parameter symmetry.

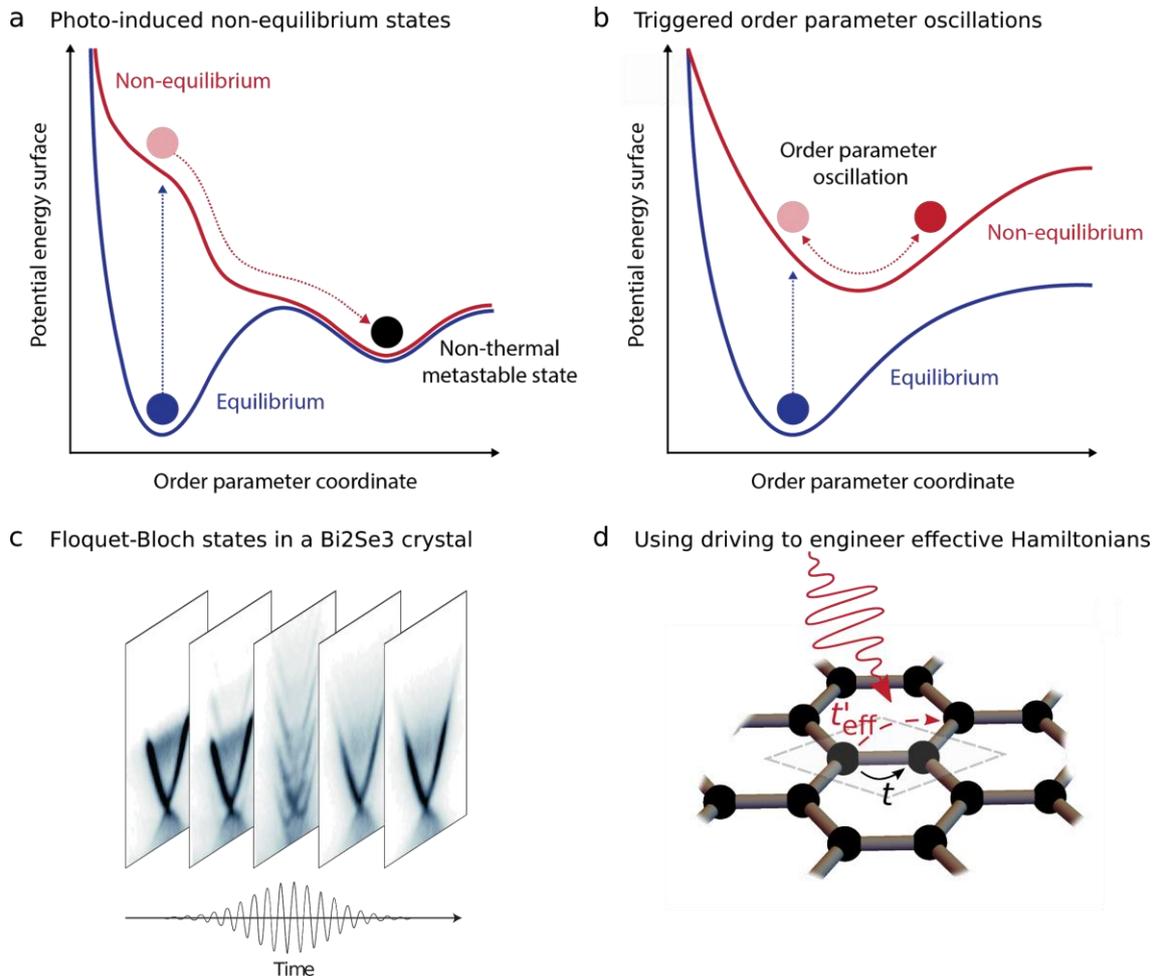

a  Photo-induced non-equilibrium states

b  Triggered order parameter oscillations

c  Floquet-Bloch states in a Bi2Se3 crystal

d  Using driving to engineer effective Hamiltonians

**Figure 1:** a) Ultrafast optical excitation can transiently modify the energy landscape of quantum materials, inducing non-equilibrium states with distinct electrical, magnetic and structural properties from those accessible in thermal equilibrium. In some cases, the system relaxes into auxiliary minima of the equilibrium energy landscape and metastability is achieved. b) When the energy landscape is only weakly perturbed by the ultrafast optical quench, oscillations of the order parameter can be initiated. c) Evolution of the Floquet-Bloch dispersion at the surface of photoexcited Bi$_2$Se$_3$ as reported in [Mahmood et al. Nature Phys. 12 , 30632 (2016)]. Note the development of the Floquet replicas. c) is reproduced from ref. [7] Macmillan Publishers Ltd. d) Driving can also be used to design effective hopping parameters (t'$_{eff}$) and Hamiltonians, as in the realization of the Haldane model in ultracold matter [32].

Much of the research to date has focused on the non-equilibrium effects induced in solids immediately following impulsive photo-excitation. In more recent years, the ability to generate strong-field optical pulses at longer wavelengths—where light-matter interaction is inherently enhanced—has led to a surge of interest in exploring phenomena that occur *during* optical illumination. Experiments are now in the regime where the material response is governed by the coherent dressing of its quantum states by the light field. Such processes can be described using Floquet theory [23-24], a non-perturbative and non-adiabatic framework that enables time-periodic Hamiltonians to be mapped to an effective static Hamiltonian. This mapping enables the highly developed theoretical infrastructure used to describe equilibrium solid-state physics to be applied to the non-equilibrium regime, which has led to some exciting predictions that are just starting to be tested.

One of the cornerstones of 'Floquet-engineering' in solids was the prediction of laser-induced band gaps in graphene [25,26], which was signalled to have an experimental sweet spot in the mid-infrared [27]. Since 2013, unique time-resolved ARPES experiments [28] demonstrated two crucial points. First, the appearance of Floquet-Bloch states—the photon-dressed electronic states in a crystal—through the visualization of a seemingly abstract band structure (Fig. 1c). Second, the opening of polarization-dependent band gaps in the dressed band structure. Beyond these spectral modifications, it was also predicted that Floquet-Bloch states can be topologically non-trivial [26,29,30] (for a review see [24]), which in Dirac systems requires breaking time-reversal symmetry through circularly polarized light [29]. These ideas were initially tested in photonic [31] and cold atom [32] (Fig 1d) experiments simulating graphene, but it was not until very recently that one of the most celebrated consequences of topological Floquet-engineering— the light-induced anomalous Hall effect [26,29,33]—was demonstrated in condensed matter [34].

## Current and Future Challenges

The overarching goals of this field are twofold. First, to understand the physics of solids away from thermal equilibrium, where quantum effects that have no equilibrium counterpart can be realized. Second, to use this knowledge to design functionalities that will have real technological impact.

This is a relatively young field and research to date has largely been that of exploration. In order for the field to mature, it is necessary to move beyond surveying the types of effects that can be induced and focus on understanding the overarching concepts that unify many of the phenomena already discovered.

Some important challenges include, but are not limited to:

- Devising materials and driving protocols for the realistic implementation of Floquet-engineering proposals in the face of dissipation, bath coupling and electronic heating [35,36]
- Improving our understanding of old problems such as shift photocurrents, though with much applied interest still lack a unified framework (see [37] and references therein)
- Improving our understanding of space-time symmetries to control driven disordered systems, either to achieve unprecedented topological phases [38], or new concepts such as time-crystals [39]
- Developing a coherent framework for dealing with new non-equilibrium topological phases including Floquet and non-Hermitian systems [40]
- Understanding the mechanisms behind photo-induced phase transitions and metastability
- Mapping non-equilibrium order parameter symmetries
- Improving our understanding of how directed lattice excitation can create new non-equilibrium phases of matter
- Turning non-equilibrium phenomena into functionalities by experimenting with devices
- Designing materials that host desirable non-equilibrium effects at higher temperatures for future applications

One path to addressing the challenges listed above is to increase feedback between experimental characterization, theoretical understanding, and materials synthesis—a successful strategy employed in many other branches of quantum materials research [1]. Experiments have long been the driving force in this field, and our theoretical understanding is steadily improving. But at the moment there is little emphasis on developing materials or heterostructures that are designed to host a given non-equilibrium phenomena when photo-excited. Doing so would allow researchers to more systematically test non-equilibrium theories and ultimately create new functionalities. Let this be a call for an era of intentional non-equilibrium materials design.

## Advances in Science and Technology to Meet Challenges

Experimentalists are now well-equipped with ultrafast light sources spanning the THz to the hard X-ray regime and a wide range of available spectroscopic probes. Many spectroscopies still suffer from poor signal-to-noise, but this will naturally alleviate as laser pulse train repetition rates continue to increase. There is also a need for ultrafast electrical and scanning probes, which are currently being developed [34, 41, 42]. Conceptually, the biggest experimental leap will occur once non-equilibrium phenomena can be engineered using cavities instead of strong laser pulses. This exciting research front is only just getting started [43-45].

On the theoretical side, the field will greatly benefit from the use of quantum simulators, as well as recent advances in time-dependent density functional theory allowing for a linear scaling with system size [46]. New learning algorithms built on these new state-of-the-art methods could also completely reshape our present simulation capabilities.

**Concluding Remarks**

Here we have given a brief overview of how coherent light-matter interaction can induce a wide range of non-equilibrium phenomena in quantum materials that could serve as functionalities in next-generation technologies. Distinctly motivated research fronts are intertwining and synergizing with other fast evolving topics such as topological phases of matter. We expect that recent technological advances, together with increased feedback between experiment- and theory-driven perspectives, will bring surprises in the coming years. Indeed, catalyzed by recent discoveries, the physics of non-equilibrium phenomena in quantum materials is becoming an even more promising *terra incognita*.

# Roadmap for 2D hyperbolic materials

Tony Low[1] and Anshuman Kumar[2]

[1]Department of Electrical and Computer Engineering,
University of Minnesota, Minneapolis, Minnesota 55455, United States
[2]Physics Department, Indian Institute of Technology Bombay, Mumbai 400076, India

## STATUS

Hyperbolic materials (HMs) are systems where two of the optical permittivity tensor components have opposite signs[1–3]. These materials can accommodate polaritonic modes due to dipole excitations of various origins, such as plasmons and phonons[4]. The constant energy surface ($\omega(\mathbf{k}) = \omega_0$) for the polariton mode in such a material takes the form of a hyperboloid (or hyperbola in 2D) in wavevector space. In contrast, most common materials have an ellipsoidal isofrequency surface for their supported polaritons. The unique hyperbolic topology of the $k-$space in these materials provides access to unprecedentedly large optical density of states, fundamentally changing the nature of light matter interaction. This property leads to a number of technologically relevant applications in the field of subwavelength imaging[5–8], biosensing[9], modulators[10], high precision internal structure diagnosis[11], thermal emission control[12] and spontaneous emission enhancement[13–15]. Traditional realizations of such HMs so far have been through artificially structured media called metamaterials or metasurfaces, where a carefully chosen geometry of constituent unit cell leads to an "effective" permittivity tensor which is hyperbolic. Apart from the constraints of nanofabrication, this approach has a fundamental limitation since the inverse of the unit cell size limits the maximum wavevector for the polariton[16, 17].

Past several years have seen the emergence of a new class of naturally occurring two dimensional (2D) van der Waals HMs, covering a hyperbolicity range from the visible all the way up to the mid IR wavelength range[4, 18]. The hyperbolic behaviour in these materials covers a range of collective excitations – plasmons, phonon polaritons and excitons, which not only opens up avenues for novel applications but also for exploring the fundamental nature of these polaritons in the presence of hyperbolicity. The first experimental report of natural 2D HMs appeared in 2014, where the hyperbolic modes were observed via scanning near field optical microscopy[19], followed by a number of other demonstrations, including far-field scattering measurements of HM nanostructures[20–23]. Multilayer hexagonal boron nitride (hBN) was shown to have a negative in-plane permittivity in the frequency range 1360–1610 cm$^{-1}$ and a negative out of plane permittivity in the range 760–825 cm$^{-1}$. Subsequent to this work, several 2D HM candidates were identified including black phosphorus[24–26], transition metal dichalcogenides[27, 28], transition metal oxides such as $\alpha-$MoO$_3$[29, 30], cuprates[31, 32], tetradymites[33, 34] and graphite[35, 36]. In Fig. 1, we present a library of these layered 2D HMs, where the spectral range and type of the hyperbolicity as well as the losses of the hyperbolic polariton mode quantified by a figure of merit FOM given by the ratio of the propagation length $(1/\Im\{k\})$ to the approximate polariton wavelength ($\sim 1/\Re\{k\}$) for the allowed wavevector $k$ is presented[37]. Among these, a particularly interesting class of 2D HMs are ones that show natural in plane hyperbolicity such as MoO$_3$ and WTe$_2$[27] and black phosphorus[24–26], although there have been some efforts towards artifically inducing in-plane hyperbolicity in other 2D materials also through nanostructuring[38].

There have been several efforts at experimentally detecting the hyperbolic nature of the polaritons in these materials, the most straightforward being through near field scanning microscopy[39, 40] and photo induced force microscopy[41–43]. In both cases, the hyperbolic polariton is launched by a nanoscale tip close to a 2D HM structure, however the former detects the light intensity whereas the latter measures the optomechanical force at the tip. The measured interference patterns reveals the wavevector at different excitation frequencies. Far field scattering measurements have also been used to detect hyperbolicity in the nanostructure of these 2D HMs, where the polarization and incidence angle dependences of the obtained scattering cross sections from these 2D HMs highlights the hyperbolic response[27, 44]. Recently, electron energy loss spectroscopy has also been used to map the hyperbolic polaritons in 2D materials[45, 46]. An electrical detection scheme has also been demonstrated[47].

## CURRENT AND FUTURE CHALLENGES

2D HMs are interesting for future applications, but several challenges remain outstanding. A critical constraint is the wafer scale growth of the 2D HMs[48] – most experimental demonstrations so far have employed mechanically exfoliated films with typical flake size on the order of tens to hundreds of microns only. Moreover, this method does not offer precise control on the number of layers of the 2D HMs. Hence, any macroscale application of 2D HMs is currently prohibitive as wafer scale platform would require uniformity in crystal orien-tation and layer numbers. There have been several efforts in the direction of large area growth of these systems, but large scale manufacturing of 2D HMs is still in its infancy[48–50]. As highlighted in Fig. 1, another fundamental challenge with these 2D HMs is the large Ohmic losses, particularly in the 2D HMs which rely on plasmons. This is a fundamental limitation related to intraband damping and the losses associated with interband transitions in these materials[25, 51]. For applications that rely on spontaneous emission enhancement near 2D HMs, these losses lead to the phenomenon of quenching of the emission[52]. Moreover, the propagation lengths of the hyperbolic polaritons become prohibitively small for any waveguiding applications[51]. A third challenge is the environmental stability of these 2D HMs[53]. For instance, black phosphorus has been shown to be unstable in ambient atmosphere which leads to deterioration in its optical response[54, 55]. Similarly it has been shown that upon room-temperature exposure to the environment for several months, TMDCs show aging effects such as cracking, changes in morphology and deteriorated optical response[56]. The fourth challenge is the active tunability of the hyperbolic modes[57, 58]. 2D HMs, particularly those in the mid-IR range, rely on phonon polaritons whose spectral region is fixed by the longitudinal and transverse optical phonon frequencies. This is a disadvantage in terms of the spectral coverage as well as the fixed dispersion of the phonon polaritons within the Reststrahlen bands[59].

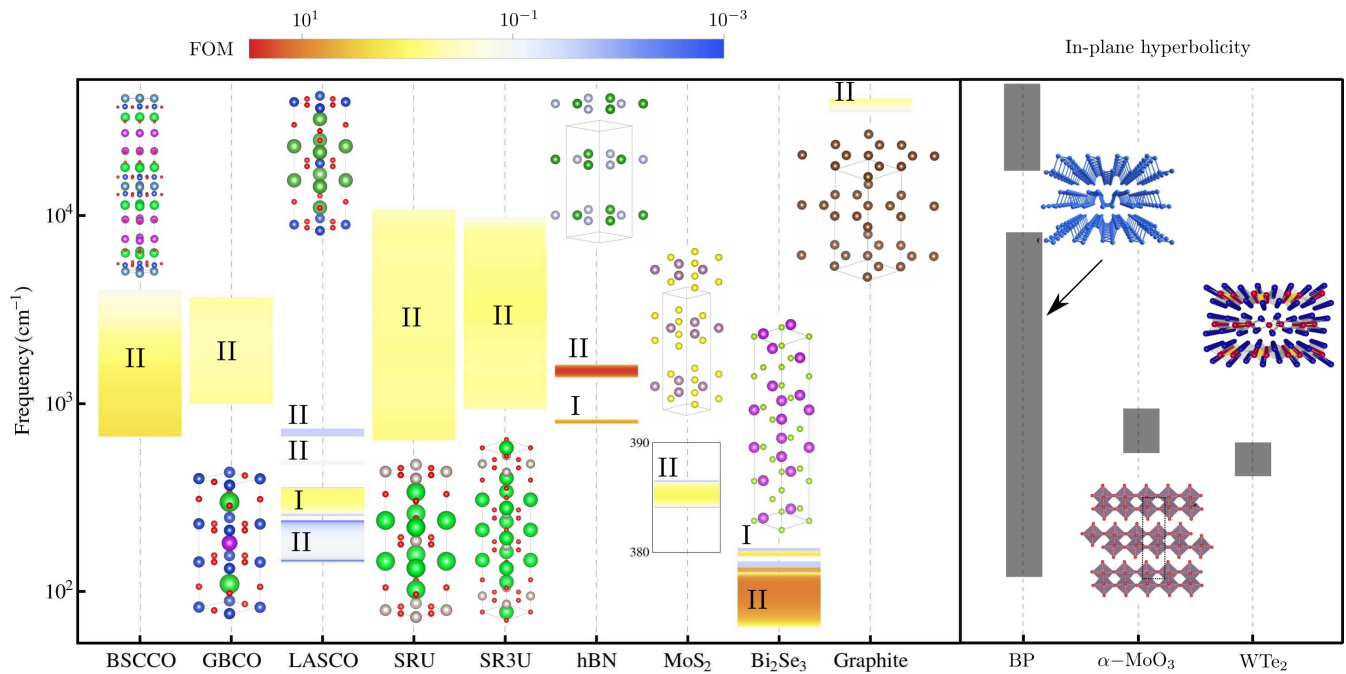

FIG. 1. Chart showing the hyperbolic frequency ranges for various naturally occurring hyperbolic layered materials, that is, cuprates (BSCCO ($Bi_2Sr_2Ca_{n_1}Cu_nO_{2n+4+x}$), GBCO ($GdBa_2Cu_3O_{7x}$), LASCO ($La_{1.92}Sr_{0.08}CuO_4$)), ruthenates(SRU ($Sr_2RuO_4$), SR3U ($Sr_3Ru_2O_7$)), hBN, TMDs, tetradymites ($Bi_2Se_3$) and graphite. The colour map depicts the calculated figure of merit (FOM) Re[q]/Im[q] for uniaxial hyperbolic materials (left part of the figure). The types of hyperbolicity I ir II is marked. Figure partially reproduced with permission from [4]. The right side of the figure depicts in plane hyperbolic materials - BP (black phosphorus) including single and multilayer with strain and doping[24–26], $\alpha-MoO_3$[29, 30] and $WTe_2$[27]. FOM for these sytems is not shown because it depends on the direction of polariton propagation in the plane of the 2D material.

## ADVANCES IN SCIENCE AND TECHNOLOGY TO MEET CHALLENGES

To meet the first challenge, namely large area growth, several recipes have been formulated[48, 49, 60]. Firstly, there have been improvements in the mechanical exfoliation itself, for instance using a gold film[61], introducing additional plasma and heat treatment steps[62]. Particularly, for TMDCs, metalorganic chemical vapor deposition has resulted in wafer scale growth[63]. For hBN, wafer scale growth has recently been demonstrated with precursors fed into molten gold [64] and Cu (111) thin films[65]. For wafer scale $\alpha-MoO_3$, a plasma enhanced chemical vapor deposition technique has been demonstrated[66]. For black phosphorus, an approach involving the conversion of red phosphorus to black phosphorus demonstrated a millimeter scale growth[5]. Despite the promise, these methods need further development for compatibility with substrates of technological interest.

Regarding overcoming the optical losses in 2D HMs,

incorporation of optical gain has been used [67, 68]. An optical gain approach was theoretically considered in [25] for black phosphorus. For hBN, the quality factor can be improved with isotopically enriched samples[69]. Another approach is working in a frequency regime where interband losses do not dominate. Here the spectral tunability of this low loss regime could be obtained by making heterostructures of 2D materials[51]. Experimental reports on both these fronts have so far been limited. The problem of instability of 2D HMs has been most severe for TMDCs and black phosphorus. Typical methods of improving on this front involve thermal treatment[53, 70] and encapsulation layers[53]. Among the latter, deposition of thin protective oxide layers via atomic layer deposition have shown promise[71, 72]. For black phosphorus in particular, hBN encapsulation has also shown promise[73–75]. A low cost encapsulation technique based on organic coatings such as PMMA[76] and PTCDA[77]. However, in the long term, such organic coatings fail to completely protect the 2D HMs from oxygen and water. As an alternative, certain chemical modification techniques are also being explored which protect the 2D HMs by covalent bonding to the surface[78, 79].

Lastly, the tunability of phonon polariton based 2D HMs is an active area of research. One possibility is the use of carrier photoinjection which has been demonstrated in other systems such as InP and 4H-SiC[59]. Tunability of hyperbolicity in black phosphorus using electrostatic gating, strain and optical pumping has been theoretically predicted[25]. There is also a recent interest in manipulating the polaritons of 2D HMs by integrating them with phase change materials such as $VO_2$[80] as well as other 2D plasmonic materials such as graphene, whose plasmons can be tuned electrostatically[8, 81]. These coupled systems then offer the possibility of active tunability of the hyperbolic polaritons and the extension of this active tunability to other 2D HMs and heterostructures presents enticing opportunities.

## CONCLUDING REMARKS

The field of 2D HMs presents several exciting open questions both from the point of view novel materials as well as improving the stability and active tunability of existing systems. Although the field is its infancy, it is rapidly growing and shows promise for many emerging applications.

---

## Spin Torque Materials

Regina Galceran[1] and Sergio O. Valenzuela[1,2]

[1]Catalan Institute of Nanoscience and Nanotechnology (ICN2), CSIC and BIST, Bellaterra, Barcelona, Spain

[2]ICREA—Institució Catalana de Recerca i Estudis Avançats, Barcelona, Spain

**Status**

In the late 90's it was discovered that a spin-polarized current can be used to manipulate the magnetization of a ferromagnet. This raised the interest not only of the spintronics community, but also of technological companies[1]. The new effect could be used in magnetic random-access memories (MRAMs), which store information in the magnetization orientation of nanoscale magnetic elements (Fig. 1a). The concept is simple: a charge current is converted into a spin current (or a non-equilibrium spin density), which in turn creates a torque on the magnetization (called "spin torque") that can switch the magnetization orientation. This approach helped overcome the scaling limitations posed by the available MRAM technology at the time, which relied on external magnetic fields.

The charge to spin (CS) conversion can be achieved by different strategies. In a first generation of commercialized spin-torque MRAMs, so called spin-transfer torque (STT) MRAM, a ferromagnetic layer is used as a polarizer in a vertical magnetic-tunnel junction (MTJ) structure (Fig. 1b)[1]. However, large current densities across the junction are required to switch the magnetization, resulting in reliability issues. A route to circumvent this problem relies on exploiting materials with large spin-orbit coupling (SOC) in contact with the magnetic element. In this case, the electrical current flows along the film stack, instead of across it, enhancing the reliability of the device. The CS-conversion process may involve bulk and interfacial SOC mechanisms [including, but not limited to, the spin Hall effect and the Rashba-Edelstein effect], opening the path for the next generation spin-orbit torques (SOT) MRAMs[2], [3]. Magnetic memories based on spin torques (such as the mentioned STT or SOT, as well as other types based on domain wall motion) are non-volatile, and offer low power consumption, high endurance, high speed, good packing density and scalability.

The following discussion is centred on SOT-MRAM. SOTs promise better efficiency and speed than STTs, and enable the switching of antiferromagnets and magnetic insulators as well as the generation of coherent spin waves[2]. However, it is currently under development and it remains a central scientific challenge to tune and engineer more efficient heterostructures. A common figure of merit and a full comprehension of the SOT phenomena in most systems are still lacking, hindering a straightforward comparison between the reported spin-torque efficiencies in different studies[2]. In parallel, new "spin torque" materials and pathways are being considered to meet the demands of increased efficiency, downscaling and switching without external magnetic fields at ultrafast speeds (Fig. 2).

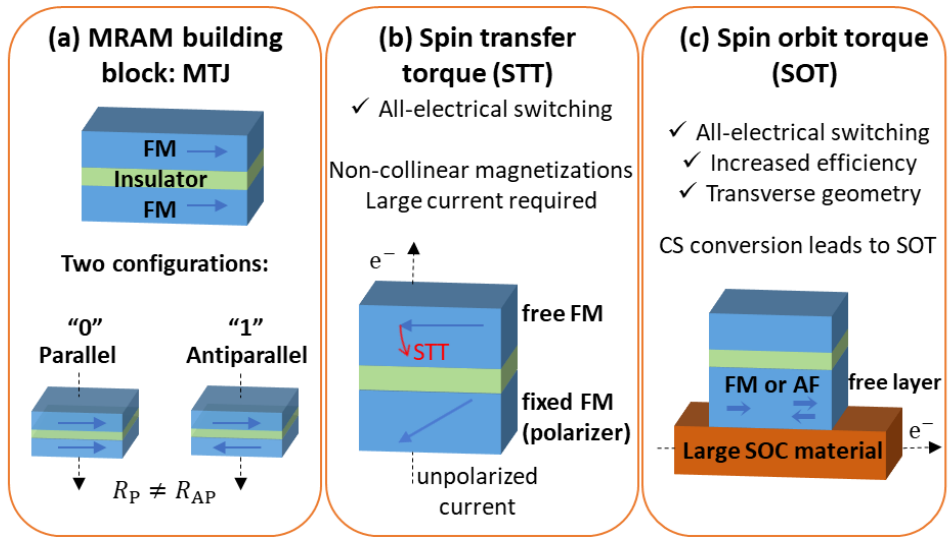

**Figure 1.** (a) Magnetic tunnel junctions are the building block of MRAM: the (parallel or antiparallel) orientation of the magnetization of two ferromagnets sandwiching a thin insulating layer defines two different states of resistance; (b) A fixed ferromagnet polarizes the current, and large (polarized) current allows the switching of the free ferromagnet by STT; (c) Current flow along a material with large spin-orbit coupling (SOC) generates a torque on an adjacent magnetic layer (ferromagnetic or antiferromagnetic), allowing its manipulation. Note that it is also possible to achieve magnetization switching by SOT in a magnetic material with large SOC, effectively combining the functionally of the two layers in one.

**Current and Future Challenges**

One major challenge is the need to reduce the current density required for magnetization manipulation and, with it, the power dissipation in devices. Using heavy metals (Pt, W...) as large SOC-materials has so far rendered a CS conversion significantly below 100%[3]. To overcome this limitation new types of materials are being investigated, including topological insulators (TI), for instance $Bi_2Se_3$ or $(Bi_{1-x}Sb_x)_2Te_3$ (Fig. 2a). TIs are insulating in the bulk but possess conducting, topologically-protected surface states in which the spin polarization is determined by the carrier propagating direction. The CS conversion efficiency in such topological surface states is potentially larger than those in Rashba split states. In fact, large SOT in TIs has been reported[4], [5], with a claimed efficiency more than one order of magnitude larger than that obtained with heavy metals, even at room temperature, while magnetization switching has already been achieved.

Parallel efforts are being made towards miniaturization of devices in which two-dimensional materials (2DM) can play an important role as we strive for ultrathin and flexible electronics (Fig. 2b). Transition metal dichalcogenides (TMDCs), which include some Weyl and Dirac semimetals, are gathering increasing attention as their spin textures could give rise to large CS conversion[6], [7]. Furthermore, the recent discovery of 2D ferromagnetic materials with metallic, e.g. $Fe_3GeTe_2$, or insulating, e.g. $CrI_3$, behaviour, open respectively the possibility of replacing the thick metal layers in conventional MTJs or creating new spin-filter MTJs that use spin-polarized tunnel barriers[8]. The reduced thickness of the overall structure with 2DMs could help overcome current limitations regarding fabrication and planarization, which are affecting the scaling of current complex MTJ pillars whose overall thickness exceeds tens of nanometres. At the same time, implementing tunnel barriers with 2DMs, such as hexagonal boron nitride (hBN) or graphene, which are atomically thin and smooth, would solve tunnel barrier uniformity issues that are arising with bulk insulators such as MgO in the ultrathin limit[9].

Another non-trivial matter is the all-electrical switching of magnetic elements with perpendicular magnetic anisotropy (PMA), which are favoured due to their better scalability for MRAM implementation (Fig. 2c). This requires torques of a certain symmetry[10], not achievable in most heterostructures (for example with heavy metals), so that switching may only be accomplished with the help of an external magnetic field. Recent reports, however, have shown that taking advantage of the reduced symmetry of the 2D semimetal WTe$_2$ or β-MoTe$_2$, torques with this type of symmetry can be obtained[6], [7]. Alternative strategies including wedged structural engineering, the use of ferroelectric substrates, domain wall pinning by geometry engineering or even taking advantage of exchange coupling, have also been considered[3].

Increasing the operation speed further motivates the study of SOT in antiferromagnetic materials[2], [11]. These are especially suited for ultrafast devices because the magnetization dynamics, governed by exchange fields, is in the THz range (whilst in ferromagnets it is limited to GHz). Antiferromagnets are being studied both as the SOT source and as the magnetic layer to be manipulated (Fig. 2d). Additionally, the lack of a macroscopic magnetic moment could help increase the integration density in magnetic memories, as well as field-free switching of PMA nano-objects by taking advantage of exchange coupling between a ferromagnet and an antiferromagnet.

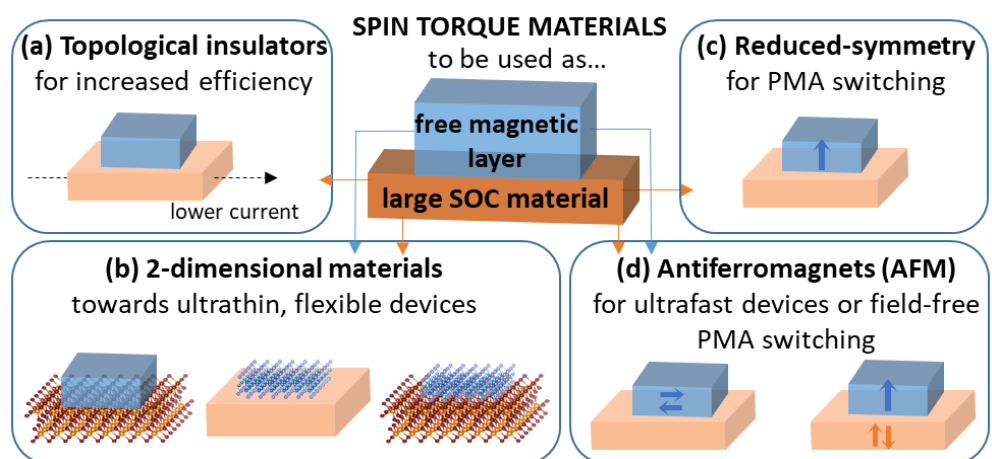

Figure 2. Schematic of new materials with potential for spin orbit torque technologies, and their possible use as large spin orbit coupling (SOC) layer (orange arrows). Additionally, 2D materials and antiferromagnets can also be embedded in the MTJ stacks (blue arrows), both as the insulating barrier or the magnetic element. The authors have used VESTA [12] and Cyrstallography Open Databases for the sketch of 2D crystal structures.

**Advances in Science and Technology to Meet Challenges**

The study of SOTs is a young research field and new types of spin torque materials are still flourishing, which accounts for the relatively few antiferromagnets, topological insulators or 2D materials that have been so far investigated. It is thus expected that many new results with novel materials will arise in the near future. Considerable progress is being made in material science towards developing high-quality heterostructures but further advances are required. They include achieving material growth optimization and large-scale integration of, for example, topological insulators, Weyl and Dirac semimetals and 2DMs with controlled thickness[10], as well as improving interfacial quality within the heterostructures' layers. Impurities (source of extrinsic contributions to the SOC), crystalline symmetries and interface oxidation and intermixing greatly influence the observed spin properties[2], [3]. Therefore, a better control of such parameters will help us advance in the reproducibility and engineering of the torques, while facilitating the understanding of the underlying physical mechanisms. Moreover, in view to consider the use of such devices in industry, further research on the compatibility of the new materials with the current Si-based technology is still pending[10].

Progress in the understanding of microscopic torque mechanisms requires the identification of the origin of large discrepancies between the reported SOT efficiencies with similar systems, which are characterized at different laboratories or with different methods[2], [5]. The wide variety of effects unrelated to SOTs that contribute to the detected signals, from Oersted fields to thermoelectric effects, illustrates the need for comparing the results obtained with alternative experimental techniques using the same heterostructure. The complex nature of the phenomena must be taken into account in the experimental design, motivating systematic measurements as a function of temperature, thickness or interlayers[2] and calling for complementary experiments with advanced characterization material science methods.

**Concluding Remarks**

SOTs represent a unique opportunity to electrically manipulate magnetic layers, enabling increased efficiency and speed towards devices such as the SOT-MRAMs exemplified above. Despite the necessary progress required both for their understanding and optimization, the potential of SOTs for applications goes well beyond simple increased MRAM performance, as it can provide novel functionalities within other technologies, for instance memristors and nanooscillators for neuromorphic computing[13].

There exists a wide variety of materials with potential in spin torque structures (for example within the families of 2D materials or topological insulators), each with distinct and unique properties. These materials are being deeply studied in other areas of condensed matter physics, offering a rich playground for interaction between fields, with much scope for yet unimaginable new applications.

**Acknowledgements**

We acknowledge support from the H2020 European Research Council PoC under Grant Agreement 899896 SOTMEM and from MINECO with grant FJCI-2016-28645.

# Magnetic skyrmions

Marius V. Costache[1] and Aurélien Manchon[2]

[1]Catalan Institute of Nanoscience and Nanotechnology (ICN2),
CSIC and BIST, Campus UAB, Bellaterra, 08193 Barcelona, Spain
[2]Aix-Marseille Université, CNRS, CINaM, Marseille, France

**Status**

Alike Majorana fermions, the concept of skyrmions originates from elementary particle physics and has inspired physicists to explain various quantum phenomena such as the quantum Hall effects. Over the past years magnetic skyrmions have become a major area of interest within the field of spintronics due to its novel physics, stimulated by the vibrant interest on topological states of matters, and its great potential for applications.

Magnetic skyrmions[1,2], bridging topology and magnetism, are local excitations in a homogeneous magnetic domain state with unique real-space topological characteristics[3]. They are part of a broader family of chiral magnetic objects that includes antiskyrmions[4], merons, bimerons etc., all featuring various forms of chiral magnetic textures. Due to the nontrivial topology of their spin texture, these skyrmions are expected to be very stable[5,6], can have a diameter of only a few nanometers, and are in principle very easy to manipulate by tiny forces[7]. Thus, they hold real potential for future low-power and high-density magnetic memory and logic applications[8,9], but also bio-inspired computing.

Magnetic skyrmions can emerge as a ground state, in which case they form a skyrmion lattice[1,2] and collective behaviors become important, or as an isolated metastable state that can be manipulated individually.

Magnetic skyrmions were first experimentally observed in bulk non-centrosymmetric crystals[1,2] and later on single ultrathin films[10]. Current approaches are largely centered on metallic heterostructures where a thin ferromagnetic layer is coupled to materials with large spin-orbit coupling. There, one can obtain skyrmions that are stable against perturbations of nanometer-scale size at room temperature. In these heterostructures, the key ingredient is the Dzyaloshinskii–Moriya magnetic interaction[11] (DMI) that emerges at interfaces upon the combination of broken inversion symmetry and strong spin-orbit interaction. Materials quality and interface control (Moriya-Dzyaloshinskii term) are key issues in skyrmions research that will allow controlling skyrmions properties in terms of size, shape, room temperature stability and mobility at low current densities.

Magnetic skyrmions can interact very efficiently with electrons and spin currents due to their non-collinear spin texture. This results in potentially high skyrmions mobility up to 1 km/s for current density of $10^8$ A/cm$^2$.

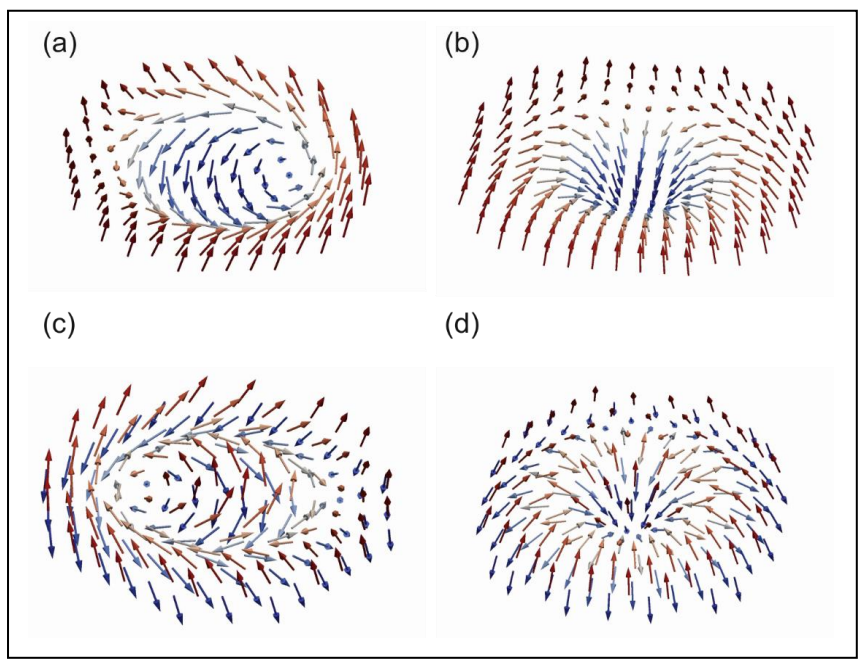

**Figure 1.** Typical skyrmion spin textures where the spin texture is indicated by the arrows. (a) Neel-type skyrmion; (b) Bloch-type skyrmion; (c) Neel-type antiferromagnetic skyrmion; (d) Bloch-type antiferromagnetic skyrmion.

## Current and Future Challenges

Although the magnetic skyrmions possess great potential for applications, there are several challenges related to materials quality and interface control, spin sensitive imaging techniques and theoretical modelling. Any utilization of skyrmions in future technological applications will require room temperature all-electrical generation, manipulation, detection and deletion of individual skyrmions.

Magnetic skyrmions can be generated by magnetic and electric fields, electric current, thermal gradients, or geometrical constrictions from domain walls. Here the challenge is to obtain better deterministic nucleation, both in space and time. The manipulation of skyrmions presents at least two challenges. One is related to the improvement of the spin-torque induce motion of skyrmions to obtain large skyrmion velocities. To do so, one can use materials with larger efficiency than heavy metals to produce spin currents such as Rashba/Dirac/valley-Zeeman interfaces. The second challenge is related to the presence of the skyrmion Hall effect that can create an undesired oblique skyrmions motion[3]. Using narrow nanowires could be an option, although skyrmions tend to vanish when reaching the wire's boundary. An alternative route is to use synthetic or real antiferromagnets, where the oblique motion is naturally suppressed[12]. There, the challenge is to control the lateral expansion occurring at high velocity[13].

For detection, recent works have focused on the electrical read-out of skyrmions by exploiting the characteristic magneto-transport (the conventional type of read-out scheme based on the TMR effect, the anomalous Hall effect), the topological Hall effect, or magnetoresistance effect measured with an STM non-magnetic tip, although the latter seems unpractical for industry.

Despite its substantial success in terms of skyrmion nucleation and manipulation, a reliable and all-electrical design integrating electric write-in, transmission and read-out of magnetic skyrmions remains a challenge. Two possible concepts exist; using the spin-transfer torque (STT) based-switching by vertical spin-polarized current injection and/or using a local electric-field effect. The use of local electric fields to create and delete individual skyrmions offers dissipation-less alternatives, compared to STT, for energy efficient skyrmion devices.

In addition to being very attractive for technological applications, magnetic skyrmions exhibit novel phenomena such as topological and skyrmion Hall effects[3]. Future research might help to understand the dynamics of skyrmions and to account for quantum and lattice effects. These rich phenomena can include Bose-Einstein condensation, localization, Mott-superfluid transition or Wigner crystallization.

**Advances in Science and Technology to Meet Challenges**

For progress to continue, substantial improvement in the understanding and control of various aspects of magnetic skyrmions is required.

For high-density and fast magnetic memory applications, very small (~nm) and high mobility (1km/s) skyrmions are necessary. Due to their nontrivial spin texture, electrons flowing through skyrmions experience a fictitious magnetic field (order of 100 Telsa) that increases upon scaling down the skyrmion and leads to topological and skyrmion Hall effects. Hence, small skyrmions are expected to display deteriorated transport properties. A better understanding of this effect is necessary to understand the edge motion of the skyrmions and identify efficient electrical detection methods.

Small and fast skyrmions require advanced engineering of ferromagnetic multilayers (to better control of the interlayer dipolar fields) but also novel structures based on synthetic ferrimagnetic and synthetic antiferromagnetic multilayers[14]. In addition, since skyrmions belong to a large family of chiral magnetic objects, the search for novel paradigms with different properties, e.g., the antiskyrmions, could open new perspectives (absence of Hall effect, better topological protection...).

Another beneficial direction is the interaction between skyrmion and thermally and electrically excited magnons. Together with a detailed study of skyrmion-skyrmion and skyrmion-anti-skyrmion interaction that could lead to the design of devices with high skyrmion mobility and novel ways of deleting/ annihilating skyrmions.

To take advantage of their huge potential for technological applications, multilayers engineering, new device design, novel materials, and advanced inspection techniques will be highly desirable. To characterize nm size skyrmions, their dynamics and their spin texture will require high-resolution microscopy techniques such as but not limited to X-ray methods, advance electron and nitrogen-vacancy center microscopy.

On the materials side, as an alternative, one can envision the use of two-dimensional materials, such as van der Waals (vdW) magnets[15], graphene, transition metal dichalcogenides (TMDC) or their combinations i.e. vdW heterostructures with several advantages. Strong spin-orbit coupling (SOC) plays a central role in these systems. For example, in vdW magnets SOC stabilizes the perpendicular magnetic anisotropy, in graphene/TMDC interfaces SOC can efficiently generate spin currents via the Rashba and Zeeman valley couplings[16]. These couplings can favour small size skyrmions with high mobility. Graphene can be used as alternative to detect skyrmions. Graphene is sensitive to the properties of the substrate on which it is placed and can inherit substrate properties due to proximity effect.

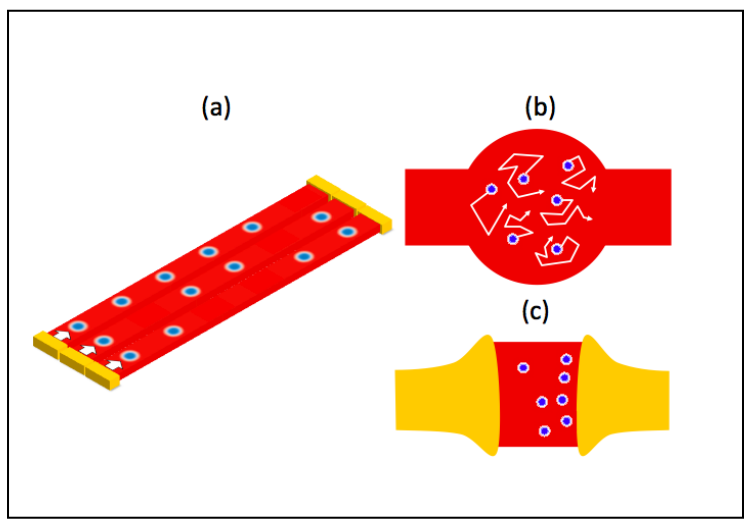

Figure 2. Schematics of three emblematic skyrmionic devices: (a) The skyrmion racetrack memory, inspired from the domain wall racetrack memory[8]. Bits of information are encoded in the skyrmions and manipulated through spin transfer torque. The working principle is similar to that of a shift register. (b) The skyrmion reshuffler realized in Ref. 9 and based on thermal diffusion of skyrmions can be used to destroy unwanted correlations in skyrmion-based probabilistic computing architecture. (c) The electrically-operated skyrmion-based artificial synaptic device can mimic the behavior of a biological synapse.

## Concluding Remarks

Magnetic skyrmions bring nm scale size, stability against external perturbations, high speed and performance advantages beyond the theoretical limits of conventional magnetic technology. Rapid advances have been achieved in materials engineering and experimental design that allowed nucleation, manipulating and deleting individual skyrmions. Novel skyrmion devices concepts have been proposed such as racetrack type memory applications, logic devices, skyrmion oscillator, bio-inspired devices, topological based communication and computing designs.

For the future, it will be necessary to substantially advance understanding and control the interaction between skyrmions and the interaction of skyrmions with spin currents and magnons. Advance in materials engineering allowing generating nm-size skyrmions that are stable at room temperature and easily manipulated with tiny currents or local electric fields together with advanced imaging methods for the study of spin texture and dynamics skyrmions would be needed. The discovery of novel chiral magnetic objects beyond skyrmions with improved properties will undoubtedly facilitate the translation towards industrial applications. Their special topological properties would allow us to explore those new physics in ways not possible with other particles. More important, given the technological importance of non-volatile magnetic memories today, the research on magnetic skyrmions will continue at least at a similar rate and level.

## Acknowledgements

M.V.C. acknowledges discussion with Nicolas Reyren and support from the Spanish Ministry of Economy and Competitiveness, MINECO (under contracts FIS2015-62641-ERC, MAT2016-75952-R and SEV-2017-0706 Severo Ochoa).

## Machine Learning Quantum Material Data

Eun-Ah Kim, Cornell University (short version)

### Status

The central goal of modern quantum materials is to search for new systems and technological paradigms that utilize quantum mechanical aspects of matter rather than being limited by them. In particular, there is an active search for new materials that exhibit surprising physical properties because of strong correlations. The key challenge in discovering these new materials lies in understanding strongly correlated quantum matter that defies both the simple theory and any ab-initio computation. Hence any progress has generally been experimentally driven. Decades of efforts in developing new experimental techniques and improving the resolution of instruments and detectors put us in an abundance of complex and comprehensive data. Two particular frontiers are (1) "image"-like real space data obtained through scanning tools and (2) high-dimensional and voluminous momentum space data obtained through scattering tools. The next breakthrough in the field will come from relating vast quantities of collected experimental data to theoretical models using tools of data science.

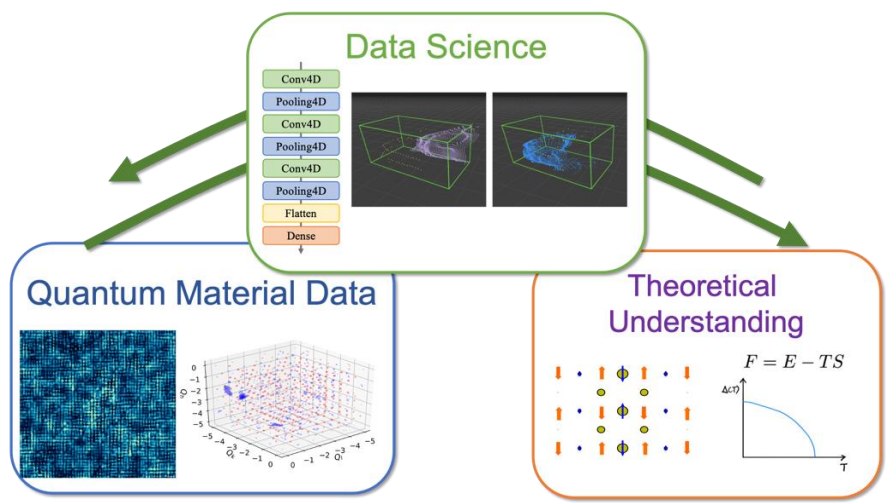

**Figure 1.** Harnessing data revolution in the quantum materials data. Data science tools can provide the critical link between theoretical understanding and complex or voluminous data now accessible through revolutionary developments in experimental techniques and detector technology.

### Current and Future Challenges

Increasingly, data science tools such as machine learning are showing great promise for accelerating discovery of materials through guided searches[1][2][3] and for improving our understanding of simulated synthetic data associated with key models[4][5][6][7]. The next challenge is to tackle the data-driven challenges forming the critical bottlenecks between comprehensive and complex experimental data on quantum materials and a theoretical understanding. Although These data-driven challenges require a fundamentally new data science approaches for two reasons: first, quantum mechanical imaging is probabilistic; and second, inference from data should be subject to fundamental laws governing microscopic interactions. Hence the ready-available data science tools need to be suitably altered to comply with hard and fast rules of fundamental microscopic laws. Moreover, any insight from learning by machines require "sanity check" or "interpretation" on machine's learning.

### Advances in Science and Technology to Meet Challenges

Recently significant first steps were taken towards matching complex experimental data such as "image"-like data to theoretical hypothesis, discriminating multiple hypothesis against each other and hence offering theoretical insight, using machine learning[89]. In Ref[8], we have developed and demonstrated a new general protocol for machine learning-based identification of symmetry-breaking ordered states in electronic-structure images. Our artificial neural networks were trained to learn the defining motifs of each category, including its topological defects, and to recognize those motifs in experimental data (Fig. 1). Applied to the hole-doped Mott insulator state harbouring the enigmatic pseudogap states[10] over a wide range of doping, the neural networks repeatedly and reliably discover the predominant features of a specific ordered state of unidirectional period $4a_0$ density wave modulations at the pseudogap energy that break the global rotational symmetry to generate a nematic state. Hence the machine learning based hypothesis testing revealed that strong coupling mechanism that drives such commensurate order is universally at play in the entire underdoped regime of the cuprate phase diagram. In Ref[9], we showed the modality developed in Ref. [8] readily extends to a completely different measurement technique of resonant ultra-sound spectroscopy.

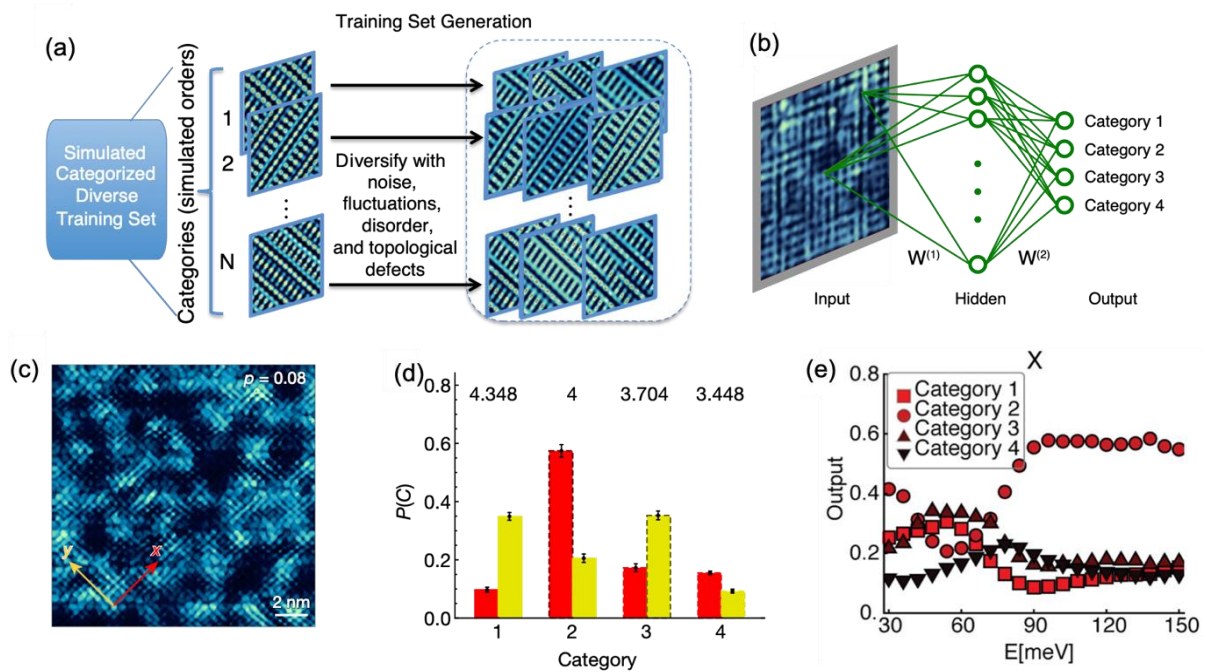

Figure 2. The machine learning based hypothesis testing on quantum material image data in practice. Figures reproduced from Ref [**Error! Bookmark not defined.**]. (a) For machine learning based hypothesis testing, ANN's are trained to recognize the given hypothesis in the synthetic data such as in electronic-structure images representing different ordered states. A training image set is synthesized by appropriately diversifying pristine images of four distinct electronic ordered states. Each translational-symmetry-breaking ordered state is labelled by a category $C = 1, 2, 3, 4$ associated with a wavelengths of $4.348a_0$, $4.000a_0$, $3.704a_0$, $3.448a_0$, respectively. The training images in each category are diversified by appropriate addition of noise, short-correlation length fluctuations in amplitude and phase, and topological defects. (b) The trained ANN is then presented with experimental data to decide among the four hypothesized states. (c) The scanning tunneling spectroscopy data of underdoped cuprate at doping level p=0.08. (d) The ANN outcome exhibits a clear preference for the commensurate period of $4.000a_0$ (category 2), especially along the X-direction. (e) A comprehensive analysis of the data set over the entire range of energies reveal the onset of unidirectional commensurate order at the pseudogap

## Concluding Remarks

The demonstrations that artificial neural networks can process and identify specific broken symmetries of highly complex experimental image arrays[8] and constrain the nature of order parameter from a sequence of peaks [9] are milestones for general scientific techniques. Moving forward, more progress in unsupervised discovery from high-dimensional and voluminous momentum space data are to be much anticipated.

**Acknowledgements**

*Eun-Ah Kim was supported by DOE basic energy sciences through Award DE-SC0018946 and by NSF, Institutes for Data-Intensive Research in Science and Engineering – Frameworks through award OAC-1934714; by the Cornell Center for Materials Research with funding from the NSF MRSEC program (DMR-1719875).*

**References**

*[(Separate from the two-page limit) Limit of 10 References. Please provide the full author list, and article title, for each reference to maintain style consistency in the combined roadmap article. Style should be consistent with all other contributions- use IEEE style]*

Machine Learning and DFT simulations of Quantum Materials

# Gabriel R Schleder [1] [2], Adalberto Fazzio [1] [2]

*¹Federal University of ABC, 09210-580, Santo André, São Paulo, Brazil*
*²Brazilian Nanotechnology National Laboratory (LNNano/CNPEM), 13083-970, Campinas, São Paulo, Brazil*

**Status**

The microscopic understanding of materials started with the paradigm shift from Classical to Quantum Mechanics in the early 20th century. Based on Schrödinger and Dirac equations, the accurate description of the behavior of materials at the fundamental level was for the first time possible. This enabled understanding of many phenomena that were not possible until then. Although in principle any problem can be described by quantum mechanics, in practice, even for simple cases this is not possible due to the complexity in solving the many-body problems. One example is superconductivity, which was discovered in 1911 but explained only 46 years later. This led to the famous Dirac remark in 1929 stating that *"the difficulty lies only in the fact that application of these laws leads to equations that are too complex to be solved".* This remark preludes the imminent central role of Computational Materials Science, presenting the main technical challenges for it to overcome.

With the impressive computational advances both in methodologies and technologies, especially since the 1970s, the description and simulation of realistic quantum materials have been possible. Density functional theory (DFT) established itself as the standard tool for simulation of materials systems after success in describing many important physical properties such as ground state structures, relative total energies, and electronic band structures. Subsequently, major milestones were achieved such as the description of structural, electronic, optical, magnetic, catalytic, and quantum properties, both for bulk and nanoscale materials. Some of these are presented in sections of this Roadmap.

Nowadays, with the ever-increasing data generation (including high-throughput techniques) and storage from sources as diverse as imaginable, and with the developments in computational and information science, a new data-driven paradigm has emerged (Fig. 1a). Building from experimental, theoretical, and computational science paradigms, the data-driven science paradigm aims to extract knowledge from this huge amount of raw data and information now available. In practice, data-driven science is embodied in novel tools and algorithms; among those, the field of statistical learning developed machine learning (ML) techniques, which are increasingly important [1]. These algorithms generate outputs given any kind of input data, whether theoretical or experimental. In materials research, many kinds of problems have been tackled, and different strategies have been thought. These can be broadly classified into learning properties of three types (Fig. 1b): *i)* pre- Schrödinger equation, such as learning the electronic density [2], to be then used as input for DFT or ML; *ii)* replacing- or accelerating- Schrödinger equation, thus creating machine-learned approximations to solving the quantum problem; and *iii)* post- Schrödinger equation, such as directly learning the outputs of the problem, as structures, energies, and materials properties. Examples include the discovery of novel 2D materials and properties [3], learning and predicting novel topological materials [4], and creating a twistronics database for identifying strongly correlated behavior in two-dimensional systems [5].

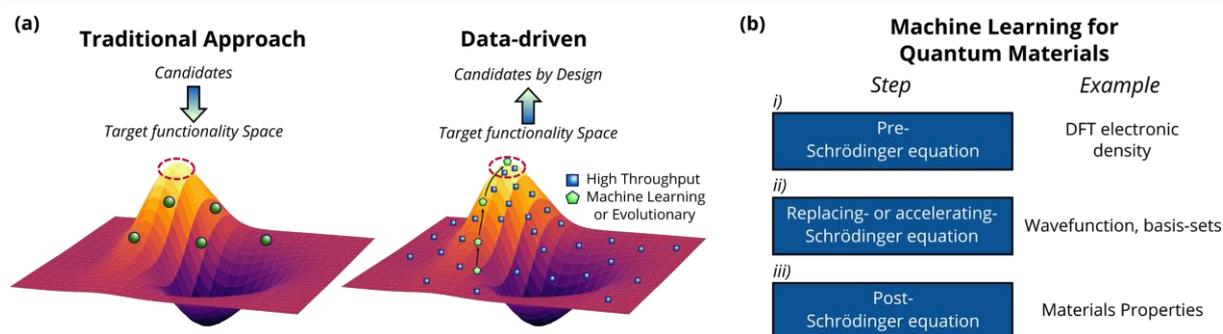

Figure 1. **(a)** Approaches in materials discovery and design: (left) traditional trial and error approach, and (right) data-driven inverse design strategies, optimizing materials target functionalities via high-throughput, machine learning, or evolutionary techniques. Reprinted with permission from [6]. Copyright 2019 American Chemical Society. **(b)** Machine learning applied to materials simulations in different stages: *i)* pre- Schrödinger equation, for instance, learning the DFT electronic density [2]; *ii)* replacing- [7] or accelerating- Schrödinger equation, e.g. Bayesian optimization for geometry relaxation, nudged-elastic band (NEB) calculations, or global optimization [8]; or *iii)* post- Schrödinger equation, such as directly learning materials properties [3].

## Current and Future Challenges

Two main materials science goals are (A) materials design/optimization (local search or exploitation), and (B) materials discovery (global search or exploration). The first aims to find among all known materials, the best candidates for each interest property or application, and then how to improve promising and interesting systems further. The second aims to discover the maximum number of novel materials, exploring simultaneously the atomic (elements), compositional (stoichiometries), and configurational (geometries/structures) spaces. As such, the number of degrees of freedom is immense, and the search (sampling) strategy in this space is of paramount importance.

Currently, both goals present major challenges. The traditional direct approach (A) includes ever-increasing complex properties such as magnetism, topology, superconductivity, and many others where quantum and many-body effects are of extreme importance. As such, on the technical side, more complex calculations demand progressively more computational power, also allowing for larger realistic systems to be treated. Challenges present for implementing sufficiently higher-level theories needed to account for quasiparticle and many-body interactions, at the same time with treatable efficiency for required system sizes. Also, developing DFT functionals better-describing correlation effects. In certain cases, the development of the fundamental theoretical framework can itself be a challenge.

In the inverse approach (B), i.e. the search for materials presenting certain desired functionalities, exploration of the large materials space is fundamental. Therefore, high-throughput, evolutionary, and machine learning techniques are increasingly important to efficiently sample the materials space, resulting in screened cases for in-depth investigation. High-throughput approaches present challenges regarding the construction of consistent databases for expensive calculations, unfeasible to be performed exhaustively. Although successful, machine learning approaches require four interdependent components (Fig. 2a), each demanding careful examination and presenting its own challenges: 1) *problem definition*, 2) *data* (*inputs*), 3) *representation*, and 4) *algorithms* and model training.

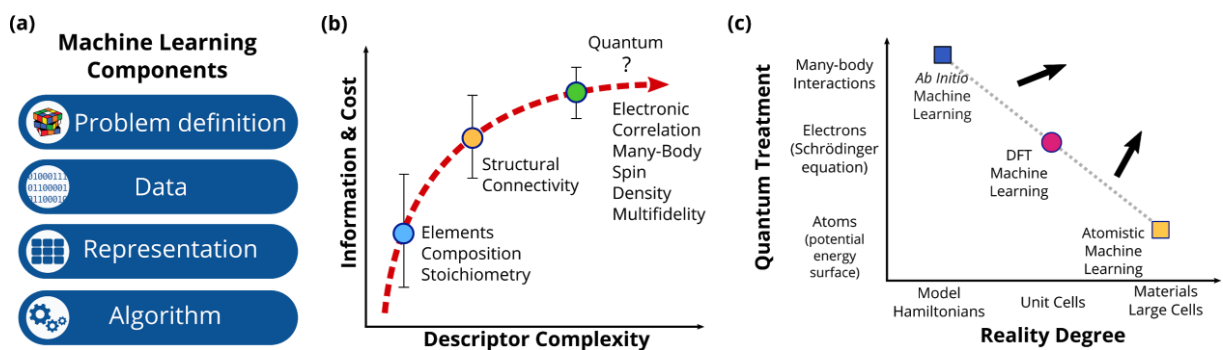

Figure 2. Machine learning for quantum materials: **(a)** 4 components. **(b)** ML representations: essential information should be added to the feature space, balancing the accuracy with efficiency. Adapted with permission from [6]. Copyright 2019 American Chemical Society. **(c)** Current ML areas for materials science. Atomistic ML allows structural exploration (global search) and properties, while neglecting quantum effects. Model Hamiltonian ML allows exploration of quantum and many-body effects, while hardly applicable to real material systems. Materials or DFT ML must bridge the gap between those areas, possible steps forward are extracting realistic Hamiltonians from DFT calculations [9] and generating wavefunctions from atomistic ML [7].

**Advances in Science and Technology to Meet Challenges**

Advances in both materials science goals mutually benefit. Beyond traditional manual developments, artificial intelligence and machine learning techniques can be broadly employed, improving every problem-solving steps. This includes estimating the computational cost, optimizing simulation and parallelization parameters, creating optimized density functionals, obtaining information for use or directly bypassing simulations, and directly obtaining results and materials properties.

Toward this, machine learning advances in each of its components are needed. First, quantum materials ML _**problem definition**_ might be supervised, i.e., regression of a numerical property or classification among different categories. If on the other hand, an exploratory and unbiased approach is more suitable, unsupervised tasks such as dimensionality reduction and clustering techniques can be preferred. Regarding _**input data**_, quantum materials generally allow for the creation of smaller databases, in contrast to big-data datasets. Accordingly, the representation used and the algorithm to be employed must also be carefully selected. Merging experimental and literature-extracted data can be valuable, with the additional benefit of improving agreement to experiments, as also using multi-fidelity, data augmentation, and transfer learning techniques. Certainly, the _**representation**_, also called _descriptor_ or _fingerprint,_ is the central element in ML for quantum materials, where nonlocal, collective, and many-body effects are essential. The challenge is distilling the fundamental information from the systems into efficient numerical descriptions (Fig 2b). Insights can be taken from two more mature ML areas (despite having larger datasets): atomistic potential energy surface (PES) ML, and many-body Hamiltonian ML (see Roadmap section #). The first usually uses additive descriptions based on local atomic environments, while the latter succeeded by translating Hamiltonians into image-like vectors or tensors commonly used with deep neural networks. Developments for quantum materials lie between these two areas, i.e. describing quantum phenomena in realistic systems, with ML as a driver for advances (Fig 2c). Lastly, the _**algorithm**_ choice must suit the problem data size and representation, for instance, deep learning can have hundreds to millions of trainable coefficients which is unreasonable for small datasets.

Finally, exciting advances are expected in the area of distilling the data-driven correlation models for driving theoretical understanding. Interpretability of generated ML models is a first step in this direction, but progress toward physics-enhanced surrogate models [3], and the broader goal of machine-assisted discovery of underlying fundamental physics [10] is essential. In practice, it is crucial to incorporate machine learning strategies into routine workflows, whether by direct integration into used software or as interoperable standalone software libraries.

## Concluding Remarks

Computational physics has traditionally been at the core of many fundamental advances related to quantum systems and materials. With the development of computational methodologies and technologies, this progress has enabled more and realistic problems to be treated with ever-increasing levels of theory. State-of-the-art DFT has driven the field forward, but it faces challenges to overcome. The expansion in information-generation capability is ideal for data-driven approaches such as machine and deep learning, which can accelerate or make possible knowledge production for the different complex problems related to quantum materials. Challenges exist related to problem definition, data acquisition, efficient and complete representations, and suitable algorithms to be used. The field is still in its beginning, and many opportunities lie ahead for future developments.

## Acknowledgements

*GRS and AF acknowledge financial support from the Fundação de Amparo à Pesquisa do Estado de São Paulo (FAPESP), project numbers 2017/18139-6 and 17/02317-2.*

## ORCID iDs

*Gabriel R. Schleder: https://orcid.org/0000-0003-3129-8682*
*Adalberto Fazzio: https://orcid.org/0000-0001-5384-7676*